\documentclass[iop,apj]{emulateapj}

\usepackage{graphicx,amsmath,amssymb, natbib}
\usepackage{multirow}
\usepackage{booktabs}
\usepackage{hyperref}

\hypersetup{
    bookmarks=true,         
    unicode=true,          
    colorlinks=true,       
    linkcolor=red,          
    citecolor=blue,        
    filecolor=magenta,      
    urlcolor=red           
}

\shorttitle{The GRB host galaxy luminosity function}
\shortauthors{Schulze et al.}

\newcommand{\hst}{\textit{HST}}
\newcommand{\spitzer}{\textit{Spitzer}}
\newcommand{\swift}{\textit{Swift}}

\begin{document}

\title{THE OPTICALLY UNBIASED GRB HOST (TOUGH) SURVEY. VII. THE HOST GALAXY LUMINOSITY
FUNCTION: PROBING THE RELATIONSHIP BETWEEN GRBs AND STAR
FORMATION TO REDSHIFT $\sim6$}

\author{S.~Schulze\altaffilmark{1,2,3},
R.~Chapman\altaffilmark{3,4},
J.~Hjorth\altaffilmark{5},
A. J.~Levan\altaffilmark{6},
P.~Jakobsson\altaffilmark{3},
G.~Bj\"ornsson\altaffilmark{3},
D. A.~Perley\altaffilmark{7,8},
T.~Kr\"uhler\altaffilmark{9},
J.~Gorosabel\altaffilmark{10,11,12},
N. R.~Tanvir\altaffilmark{13},
A.~de Ugarte Postigo\altaffilmark{10,5},
J. P. U.~Fynbo\altaffilmark{5},
B. Milvang-Jensen\altaffilmark{5},
P.~M\o{}ller\altaffilmark{14},
D. J.~Watson\altaffilmark{5}
}

\altaffiltext{1}{Instituto de Astrof\'isica, Facultad de F\'isica, Pontificia Universidad Cat\'olica de Chile, Vicu\~{n}a Mackenna 4860, 7820436 Macul, Santiago, Chile; sschulze@astro.puc.cl} 
\altaffiltext{2}{Millennium Institute of Astrophysics, Vicu\~{n}a Mackenna 4860, 7820436 Macul, Santiago, Chile} 
\altaffiltext{3}{Centre for Astrophysics and Cosmology, Science Institute, University of Iceland, Dunhagi 5, 107 Reykjav\'ik, Iceland} 
\altaffiltext{4}{Centre for Astrophysics Research, University of Hertfordshire, Hatfield, Herts AL10 9AB, UK} 
\altaffiltext{5}{Dark Cosmology Centre, Niels Bohr Institute, University of Copenhagen, Juliane Maries Vej 30, DK-2100 Copenhagen \O, Denmark} 
\altaffiltext{6}{Department of Physics, University of Warwick, Coventry, CV4 7AL, UK} 
\altaffiltext{7}{Department of Astronomy, California and of Technology, MC 249-17, 1200 East California Blvd, Pasadena CA 91125, USA} 
\altaffiltext{8}{Hubble Fellow} 
\altaffiltext{9}{European Southern Observatory, Alonso de C\'{o}rdova 3107, Vitacura Casilla 19001, Santiago, Chile} 
\altaffiltext{10}{Instituto de Astrofísica de Andaluc\'ia, Consejo Superior de Investigaciones Cient\'ificas (IAA-CSIC), Glorieta de la Astronom\'ia s/n, E-18008 Granada, Spain} 
\altaffiltext{11}{Unidad Asociada Grupo Ciencias Planetarias UPV/EHU-IAA/CSIC, Departamento de F\'{\i}sica Aplicada I, E.T.S. Ingenier\'{\i}a, Universidad del Pa\'{\i}s Vasco UPV/EHU, Alameda de Urquijo s/n, E-48013 Bilbao, Spain} 
\altaffiltext{12}{Ikerbasque, Basque Foundation for Science, Alameda de Urquijo 36-5, E-48008 Bilbao, Spain} 
\altaffiltext{13}{Department of Physics and Astronomy, University of Leicester, University Road, Leicester LE1 7RH, UK} 
\altaffiltext{14}{European Southern Observatory, Karl-Schwarzschildstrasse 2, D-85748, Garching, Germany} 

\begin{abstract}
Gamma-ray bursts (GRBs) offer a route to characterizing star-forming
galaxies and quantifying high-$z$ star formation that is distinct from the
approach of traditional galaxy surveys: GRB selection is independent of dust
and probes even the faintest galaxies that can evade detection in flux-limited
surveys. However, the exact relation between the GRB rate and the star formation rate
(SFR) throughout all redshifts is controversial. The Optically Unbiased GRB Host (TOUGH)
survey includes observations
of all GRB hosts (69) in an optically unbiased sample of \swift\ GRBs and we utilize
these to constrain the evolution of the UV GRB-host-galaxy luminosity function (LF)
between $z=0$ and $z=4.5$, and compare this with LFs derived from both
Lyman-break galaxy (LBG) surveys and simulation modeling. At all redshifts we find
the GRB hosts to be most consistent with a luminosity function derived
from SFR weighted models incorporating GRB production via both
metallicity-dependent and independent channels with a relatively high level of
bias toward low metallicity hosts. In the range $1<z<3$ an SFR weighted
LBG derived (i.e., non-metallicity biased) LF is also a reasonable fit to the data.
Between $z\sim3$ and $z\sim6$, we observe an apparent lack of UV bright hosts in
comparison with LBGs, though the significance of this shortfall is
limited by nine hosts of unknown redshift. 

\end{abstract}


\received{2015 March 13} \accepted{2015 June 3}

\section{Introduction}

The past decade in extragalactic astronomy has been characterized by a tremendous
increase in the understanding of the properties of high-$z$ galaxies \citep[for
reviews see e.g.,][]{Wolfe2005a, Shapley2011a, Carilli2013a}, such as the
diversity and evolution of their star formation rate (SFR) histories
\citep{Pettini2002a, Hopkins2004a, Hopkins2006a, Bouwens2007a, Bouwens2010a},
the relation between stellar mass and luminosity \citep{Pettini2001a, Magdis2010a,
Schaerer2013a}, the relation between mass and metallicity
\citep[e.g.,][]{Tremonti2004a, Savaglio2005a, Erb2006a, Lee2006a, Foster2012a}
and the nature of extinction as a function of luminosity \citep{Meurer1999a,
Calzetti2000a}. This progress has been driven by multi-band flux-limited surveys
of ever-improving depth and coverage permitting the
study of galaxies at redshifts ranging from $z = 0$--10 \citep[e.g.,][]{Scoville2007a,
Bouwens2014b}.

These surveys are however not suited to assess the contribution of the faintest
galaxies to the cosmic star formation history. Gamma-ray burst (GRB) selected
galaxy studies provide a complementary approach to constrain galaxy evolution
across the whole mass spectrum \citep{Perley2009a, Kruehler2011a, Rossi2012a}
and from very low to very high redshift (the most distant GRB with a spectroscopic
redshift known to date is GRB 090423 at $z=8.2$; \citealt{Tanvir2009a,
Salvaterra2009a}). The advantage of GRB-selected galaxy studies is that GRB
production requires (in its simplest form) only a massive star \citep[for a
review see][and references therein]{Woosley2011a}, which makes their detection
independent of galaxy luminosity.

The exact relation between the GRB rate and the SFR is controversial: while long-duration
GRBs are produced by massive stars and sample the entire range of known star-forming
galaxies from faint dwarfs up to luminous Lyman-break galaxies (LBGs;
\citealt{Steidel1996a}) and sub-millimeter galaxies \citep[e.g.,][]{Christensen2004a,
Tanvir2004a, Fruchter2006a, Michalowski2008a, Savaglio2009a, Kruehler2011a, Rossi2012a, Perley2013a,
Hunt2014a,Schady2014a,Kohn2015a}, most of our knowledge is based on heterogeneous samples. Several GRB
luminosity function (LF) and redshift distribution studies \citep[e.g.,][]{
Kistler2008a, Jakobsson2012a, Robertson2012a} have found that the numbers of
GRBs produced at high redshift imply that either the global SFR density is
greater at high redshift than found from LBG surveys, or that GRB production
efficiency increases with redshift. Other authors consider that this result can
be explained by continuing observational and redshift biases in existing GRB
surveys \citep[e.g.,][]{Elliott2012a}. In this work, we approach this question
from the unique standpoint of having observations of the properties of the
individual hosts of each burst in an observationally unbiased GRB sample, thus
providing a complementary approach to simulation studies.

``The Optically Unbiased GRB Host" (TOUGH) survey by \citet{Hjorth2012a} is the
first such survey to make use of the strategic advantage of \swift\ to realize
the production of a GRB host galaxy sample selected solely by accurate X-ray
localization and VLT observability, and unbiased by optical criteria such as
afterglow detection or brightness. In this paper, we compare the
evolution of the UV LF of GRB host galaxies to those derived
both from LBG samples and those predicted from stellar population synthesis
models \citep{Trenti2014a} which include both metallicity dependent and
independent channels for GRB production.

Throughout the paper, we assume a $\Lambda$CDM cosmology with
$H_0 = 71~\rm{km\,s}^{-1}\, \rm{Mpc}^{-1}$, $\Omega_{\rm m} = 0.27$, and
$\Omega_{\Lambda} = 0.73$ \citep{Larson2011a}. All reported magnitudes are
given in the AB system, and uncertainties are given at $1\sigma$ confidence
level, if not stated otherwise.

\section{Observations and Data Reduction}\label{sec:data_selection}
\subsection{The TOUGH Survey}

The TOUGH survey targeted 69 GRBs in the $R$ and $K_s$ bands with FORS2 and
ISAAC reaching limiting magnitudes of $R({\rm AB })\sim27.3$ mag and
$K_s({\rm AB})\sim23.4$ mag at $3\sigma$ confidence level. About 80\%
have a detected host galaxy \citep[][see also Malesani et al. 2015, in preparation]{Hjorth2012a}
and thanks to ongoing spectroscopic follow-up observations $>87\%$ now have a
measured redshift (\citealt{Jakobsson2012a, Kruehler2012a, Kruehler2015a}).
In addition, the hosts between $z=2$ and $z=4.5$ were targets of a
moderately deep spectroscopy campaign to study Ly$\alpha$ in emission, and hosts
at $z<1$ were part of a radio survey. These campaigns allowed the investigation
of several properties such as the $(R-K_s)$-color and the offset distribution
\citep[][Malesani et al. 2015, in preparation]{Hjorth2012a}, the redshift distribution
\citep{Jakobsson2012a}, the Ly$\alpha$ recovery rate \citep{MilvangJensen2012a},
and the unobscured SFR inferred from radio observations
for the hosts at $z<1$ \citep[][]{Michalowski2012a}.

\subsection{New Data and Their Reduction}\label{sec:data_red}

\begin{table}
\caption{Log of GRB Host Observations}
\centering
\scriptsize
\begin{tabular}{llccccc}
\toprule
\multicolumn{1}{c}{\multirow{2}{*}{GRB}}	& \multicolumn{1}{c}{Telescope}	& \multicolumn{1}{c}{\multirow{2}{*}{Filter}}	& \multicolumn{1}{c}{\multirow{2}{*}{Date}}	& \multicolumn{1}{c}{Exposure}\\
											& \multicolumn{1}{c}{Instrument}&												&											& \multicolumn{1}{c}{Time (s)}\\
\midrule
050525A		& GTC/OSIRIS	& $g'$		& 2012 Aug 22	& $24\times240$		\\
050730		& \hst/ACS	& $F775W$	& 2010 Jun 10	& $7844$		\\
050803		& GTC/OSIRIS	& $g'$		& 2014 Jul 21	& $6\times360$		\\
050803		& Keck/LRIS	& $g'$		& 2011 Aug 28	& $4\times200$		\\
050803		& Keck/LRIS	& $R$		& 2011 Aug 28	& $4\times170$		\\
050803		& Keck/ESI	& $R$		& 2005 Aug 04	& $2\times180$		\\
050803		& GTC/OSIRIS	& $i'$		& 2014 Jul 21	& $15\times120$		\\
050803		& \hst/WFC3	& $F160W$	& 2011 Sep 03	& 906			\\
050803		& \spitzer/IRAC	& $3.6~\mu\rm m$& 2013 Jan 31	& $54\times100$		\\
050803		& \spitzer/IRAC	& $4.5~\mu\rm m$& 2013 Jan 31	& $54\times100$		\\
050824		& TNG/LRS	& $B$		& 2010 Oct 13	& $2\times900$		\\
051016B		& Gemini-S/GMOS	& $g'$		& 2014 Feb 07	& $4\times100$		\\
051117B		& Gemini-S/GMOS	& $u'$		& 2014 Jan 30	& $9\times60$		\\
050904		& \hst/ACS	& $F850LP$	& 2005 Sep 26	& 4216			\\
050908		& \hst/ACS	& $F775W$	& 2010 Oct 31	& 7892			\\
050922B		& Keck/LRIS	& $g'$		& 2008 Aug 03	& $2\times360$		\\
050922B		& GTC/OSIRIS	& $i'$		& 2014 Jul 22	& $20\times120$		\\
050922B		& GTC/OSIRIS	& $z'$		& 2014 Jul 22	& $30\times60$		\\
050922B		& \spitzer/IRAC	& $3.6~\mu\rm m$& 2013 Aug 30	& $54\times100$		\\
050922B		& \spitzer/IRAC	& $4.5~\mu\rm m$& 2013 Aug 30	& $54\times100$		\\
060115		& \hst/ACS	& $F814W$	& 2010 Aug 27	& 7910			\\
060218		& SDSS		& $u'$		& 2004 Sep 21	& \nodata		\\
060522		& \hst/ACS	& $F110W$	& 2010 Oct 17	& 8395			\\
060526		& \hst/ACS	& $F775W$	& 2009 Aug 09	& 7844			\\
060605		& \hst/ACS	& $F775W$	& 2010 Oct 06	& 7862			\\
060607A		& \hst/ACS	& $F775W$	& 2010 Sep 17	& 7910			\\
060729		& Gemini-S/GMOS	& $g'$		& 2008 Jan 29	& $15\times180$		\\
060805A		& Keck/LRIS	& $g'$		& 2008 Feb 12	& 1080			\\
060912		& TNG/LRS	& $B$		& 2010 Oct 13	& $2\times150$		\\
060927		& \hst/WFC3	& $F110W$	& 2010 Sep 25	& 13992			\\
061021		& Keck/LRIS	& $g'$		& 2007 Dec 13	& 560			\\
061110A$^{\rm a}$& Keck/LRIS	& $V$		& 2006 Nov 21	& 680			\\
061110B		& \hst/ACS	& $F775W$	& 2010 Sep 23	& 7862			\\
070721B		& \hst/ACS	& $F775W$	& 2010 Nov 13	& 7844			\\
\bottomrule
\end{tabular}
\tablecomments{}
\tablenotetext{1}{The observation was performed 21 days after the GRB.
The afterglow was very faint so that only the accompanying GRB-SN could contaminate
the host measurement. However, GRB-SNe have a red spectrum and the observed bandpass
probed the rest-frame $u'$ band, which makes any contamination unlikely.
}
\label{tab:obs_log}
\end{table}

\begin{table*}
\caption{UV Properties of the TOUGH Sample}
\centering
\scriptsize
\begin{tabular}{l@{\hspace{1mm}}c@{\hspace{1mm}}c@{\hspace{1mm}}c@{\hspace{1mm}}c@{\hspace{1mm}}c@{\hspace{5mm}}l@{\hspace{1mm}}c@{\hspace{1mm}}c@{\hspace{1mm}}c@{\hspace{1mm}}c@{\hspace{1mm}}c}
\toprule
\multicolumn{1}{l}{\multirow{2}*{GRB}}	& \multicolumn{1}{c}{\multirow{2}*{Redshift$^{\rm a}$}}	& \multicolumn{1}{c}{\multirow{2}*{Filter}}	& $m$	& \multicolumn{1}{c}{\multirow{2}*{$\beta_{\textrm{UV}}$}}	& $M_{1600\,\textrm{\AA}}$$^{\rm b}$					& \multicolumn{1}{l}{\multirow{2}*{GRB}}	& \multicolumn{1}{c}{\multirow{2}*{Redshift}}	& \multicolumn{1}{c}{\multirow{2}*{Filter}}	& $m$	& \multicolumn{1}{c}{\multirow{2}*{$\beta_{\textrm{UV}}$}}	& $M_{1600\,\textrm{\AA}}$$^{\rm b}$\\
										&												&											& (mag)	&															& (mag)										& 										&												&											& (mag)	&															& (mag)										\\
\midrule
\multicolumn{12}{c}{\textbf{Hosts with Known Redshifts}}\\
\midrule
050315		&$ 1.95			$&$	R	$&$	24.51\pm0.15	     $&$ -1.54\mp0.03	$&$ -20.08\pm0.14		$& 060707		&$ 3.42			$&$	R	$&$ 25.01\pm0.06	$&$ -1.62\mp0.01		$&$ -20.82\pm0.06	$\\
050318		&$ 1.444		$&$	R	$&$	>26.95		     $&$ <-1.92		$&$ >-17.16			$& 060708		&$ 1.92\pm0.12		$&$	R	$&$ 26.94\pm0.28	$&$ -1.89\mp0.03		$&$ -17.75\pm0.27	$\\
050401		&$ 2.898		$&$	R	$&$	26.19\pm0.31	     $&$ -1.79\mp0.04	$&$ -19.31\pm0.31		$& 060714		&$ 2.71			$&$	R	$&$ 26.46\pm0.28	$&$ -1.82^{-0.03}_{+0.04}	$&$ -18.91\pm0.28	$\\
050406		&$ 2.7^{+0.29}_{-0.41}	$&$	R	$&$	26.76\pm0.34	     $&$ -1.86\mp0.04	$&$ -18.60\pm0.34		$& 060719		&$ 1.53			$&$	R	$&$ 24.81\pm0.12	$&$ -1.64\mp0.02		$&$ -19.27\pm0.11	$\\
050416A		&$ 0.653		$&$	g'	$&$	24.00\pm0.03	     $&$ -1.73\mp0.01	$&$ -18.26\pm0.03		$& 060729		&$ 0.54			$&$	g'	$&$ 25.30\pm0.16	$&$ -1.92\mp0.02		$&$ -16.66\pm0.15	$\\
050502B$^{\rm c}$&$ 5.2\pm0.3		$&$	R	$&$	>25.98		     $&$ <-1.85		$&$ >-20.61			$& 060805A$^{\rm e}$	&$ 0.60			$&$	g'	$&$ 24.61\pm0.07	$&$ -1.83\mp0.01		$&$ -17.53\pm0.06	$\\
050525A		&$ 0.606		$&$	g'	$&$	>25.83		     $&$ <-1.95		$&$ >-16.40			$& 060805A$^{\rm e}$	&$ 2.36			$&$	R	$&$ 25.26\pm0.14	$&$ -1.65\mp0.02		$&$ -19.79\pm0.13	$\\
050714B		&$ 2.438		$&$	R	$&$	25.51\pm0.20         $&$ -1.69\mp0.03	$&$ -19.61\pm0.19		$& 060814		&$ 1.92			$&$	R	$&$ 22.96\pm0.11	$&$ -1.19\mp0.03		$&$ -21.46\pm0.10	$\\
050730		&$ 3.969		$&$	F775W	$&$	>27.50		     $&$ <-1.98		$&$ >-18.58			$& 060908		&$ 1.88			$&$	R	$&$ 25.66\pm0.18	$&$ -1.74\mp0.03		$&$ -18.93\pm0.17	$\\
050801		&$ 1.38\pm0.07		$&$	R	$&$	>26.74		     $&$ <-1.91		$&$ >-17.26			$& 060912A		&$ 0.94			$&$	B	$&$ 22.59\pm0.06	$&$ -1.33^{-0.01}_{+0.02}	$&$ -20.39\pm0.06	$\\
050819		&$ 2.504		$&$	R	$&$	23.99\pm0.09         $&$ -1.39\mp0.02	$&$ -21.13\pm0.09		$& 060923A$^{\rm f}$	&$ 2.47^{+0.33}_{-0.52}	$&$	R	$&$ 26.12\pm0.24	$&$ -1.78\mp0.03		$&$ -19.05\pm0.23	$\\
050820A		&$ 2.615		$&$	F775W	$&$	26.30\pm0.06         $&$ -1.80\mp0.01	$&$ -18.96\pm0.06		$& 060923B		&$ 1.51			$&$	R	$&$ 24.15\pm0.16	$&$ -1.53\mp0.03		$&$ -19.83\pm0.14	$\\
050822		&$ 1.434		$&$	R	$&$	24.36\pm0.08         $&$ -1.59\mp0.01	$&$ -19.54\pm0.07		$& 060927		&$ 5.46			$&$	F110W	$&$ >28.23		$&$ <-2.10			$&$ >-18.39		$\\
050824		&$ 0.828		$&$	B	$&$	24.11\pm0.16         $&$ -1.67\mp0.03	$&$ -18.71\pm0.15		$& 061007		&$ 1.26			$&$	R	$&$ 24.56\pm0.17	$&$ -1.65\mp0.03		$&$ -19.08\pm0.15	$\\
050904		&$ 6.295		$&$	F850LP	$&$	>26.27		     $&$ <-1.95		$&$ >-20.58			$& 061021		&$ 0.35			$&$	g'	$&$ 26.40\pm0.16	$&$ -2.07\mp0.01		$&$ -14.65\pm0.15	$\\
050908		&$ 3.347		$&$	F775W	$&$	27.55^{+0.30}_{-0.24}$&$ -1.96\mp0.03	$&$ -18.22^{+0.30}_{-0.24}	$& 061110A		&$ 0.76			$&$	V	$&$ 25.25\pm0.10	$&$ -1.85\mp0.01		$&$ -17.41\pm0.09	$\\
050915A		&$ 2.527		$&$	R	$&$	24.70\pm0.16	     $&$ -1.54\mp0.03	$&$ -20.47\pm0.15		$& 061110B		&$ 3.43			$&$	F775W	$&$ 27.02^{+0.26}_{-0.21}$&$ -1.91^{-0.03}_{+0.02}	$&$ -18.79^{+0.26}_{-0.21}$\\
050922C		&$ 2.199		$&$	R	$&$	>26.29		     $&$ <-1.81		$&$ >-18.64			$& 061121		&$ 1.32			$&$	R	$&$ 22.84\pm0.03	$&$ -1.31\mp0.01		$&$ -20.67\pm0.03	$\\
051001		&$ 2.43			$&$	R	$&$	24.53\pm0.13	     $&$ -1.51\mp0.03	$&$ -20.55\pm0.12		$& 070103		&$ 2.62			$&$	R	$&$ 24.21\pm0.14	$&$ -1.43\mp0.03		$&$ -21.03^{+0.13}_{-0.14}$\\
051006		&$ 1.059		$&$	R	$&$	23.03\pm0.07	     $&$ -1.43\mp0.01	$&$ -20.03\pm0.06		$& 070110		&$ 2.35			$&$	R	$&$ 25.19\pm0.11	$&$ -1.64\mp0.02		$&$ -19.84\pm0.11	$\\
051016B		&$ 0.936		$&$	g'	$&$	23.13\pm0.03	     $&$ -1.46\mp0.01	$&$ -19.87\pm0.03		$& 070129		&$ 2.34			$&$	R	$&$ 24.23\pm0.12	$&$ -1.45\mp0.03		$&$ -20.75\pm0.11	$\\
051117B		&$ 0.481		$&$	u'	$&$	22.91\pm0.19	     $&$ -1.65\mp0.03	$&$ -18.67\pm0.18		$& 070224		&$ 1.99			$&$	R	$&$ 26.02\pm0.31	$&$ -1.78\mp0.04		$&$ -18.71^{+0.30}_{-0.29}$\\
060115		&$ 3.533		$&$	F814W	$&$	27.21^{+0.27}_{-0.21}$&$ -1.94\mp0.03	$&$ -18.65\pm0.27		$& 070306		&$ 1.50			$&$	g'	$&$ 22.90\pm0.09	$&$ -1.18\mp0.03		$&$ -21.21\pm0.09	$\\
060218		&$ 0.034		$&$	u'	$&$	20.61\pm0.12	     $&$ -2.02\mp0.01	$&$ -15.20\pm0.11		$& 070318		&$ 0.84			$&$	R	$&$ 24.60\pm0.11	$&$ -1.76\mp0.01		$&$ -18.18\pm0.10	$\\
060306$^{\rm d}$&$ 1.559		$&$	R	$&$	24.21\pm0.08	     $&$ -1.54\mp0.02	$&$ -19.86\pm0.07		$& 070328		&$ 2.06			$&$	R	$&$ 24.55\pm0.13	$&$ -1.54\mp0.03		$&$ -20.17\pm0.12	$\\
060522		&$ 5.11			$&$	F110W	$&$	>27.82		     $&$ <-2.05		$&$ >-18.69			$& 070419B		&$ 1.96			$&$	R	$&$ 25.20\pm0.20	$&$ -1.66\mp0.03		$&$ -19.44\pm0.19	$\\
060526		&$ 3.221		$&$	F775W	$&$	>27.52	   	     $&$ <-1.95		$&$ >-18.18			$& 070506		&$ 2.31			$&$	R	$&$ 26.21\pm0.22	$&$ -1.79\mp0.03		$&$ -18.83\pm0.21	$\\
060604		&$ 2.136		$&$	R	$&$	25.62\pm0.18	     $&$ -1.71\mp0.03	$&$ -19.23\pm0.17		$& 070611		&$ 2.04			$&$	R	$&$ >27.27		$&$ <-1.92			$&$ >-17.55		$\\
060605		&$ 3.773		$&$	F775W	$&$	27.48^{+0.40}_{-0.29}$&$ -1.97\mp0.03	$&$ -18.50^{+0.40}_{-0.29}	$& 070721B		&$ 3.63			$&$	F775W	$&$ 27.69^{+0.47}_{-0.33}$&$ -1.99^{-0.04}_{+0.03}	$&$ -18.22^{+0.47}_{-0.33}$\\
060607A		&$ 3.075		$&$	R	$&$	>28.05	             $&$ <-1.99		$&$ >-17.57			$& 070802		&$ 2.45			$&$	R	$&$ 25.25\pm0.21	$&$ -1.64\mp0.04		$&$ -19.88\pm0.20	$\\
060614		&$ 0.125		$&$	u	$&$	24.71\pm0.30	     $&$ -2.10\mp0.01	$&$ -14.06\pm0.29		$\\																		
\midrule
\multicolumn{12}{c}{\textbf{Hosts with Unknown Redshifts$^{\rm g}$}}\\
\midrule
050726		& 3.5			&$	R	$&$	>26.19       	     $&$ <-1.81			$&$ >-19.68		$& 061004		& 3.5			&$	R	$&$ >25.84		$&$ <-1.76			$&$ >-20.03		$\\
050803		& 3.5			&$	i'	$&$	26.29\pm0.50	     $&$ -1.83^{-0.06}_{+0.07}	$&$ -19.55\pm0.50	$& 070330		& 3.5			&$	R	$&$ >26.19		$&$ <-1.81			$&$ >-19.67		$\\
050922B		& 3.5			&$	i'	$&$	25.20\pm0.15	     $&$ -1.67\mp0.03		$&$ -20.63\pm0.15	$& 070621		& 3.5			&$	R	$&$ 25.85\pm0.23	$&$ -1.77^{-0.03}_{+0.04}	$&$ -20.02\pm0.23	$\\
060919		& 3.5			&$	R	$&$	25.80\pm0.26	     $&$ -1.76\mp0.04		$&$ -20.07\pm0.26	$& 070808		& 3.5			&$	R	$&$ 26.85\pm0.33	$&$ -1.90\pm0.04		$&$ -19.01\pm0.33	$\\
060923C		& 3.5			&$	R	$&$	25.49\pm0.18	     $&$ -1.71\mp0.03		$&$ -20.38\pm0.18	$\\																												
\midrule
\multicolumn{12}{c}{\textbf{New Photometric Redshifts$^{\rm h}$}}\\
\midrule
050803		&$3.5\pm0.5		$&$	i'	$&$	26.29\pm0.5	     $&$ -1.83^{-0.06}_{+0.07}	$&$ -19.55\pm0.50	$& 050922B 		&$ 4.5\pm0.5		$&$	i'	$&$ 25.20\pm0.15	$&$ -1.69\mp0.03		$&$ -21.13\pm0.15	$\\
\bottomrule
\end{tabular}
\tablecomments{All magnitudes are corrected for Galactic extinction
		and were converted into the AB system. For the non-detections we report the
		$3\sigma$ limiting magnitudes. Redshifts were taken from \citet{Hjorth2012a},
		\citet{Perley2013a} and \citet{Kruehler2015a}.}
\tablenotetext{1}{For hosts with photometric redshifts, we report the values at
		the nominal redshifts.}
\tablenotetext{2}{The absolute magnitude was computed through
		$M_{1600\,\textrm{\AA}}=m-{\rm{DM}}(z)-2.5\,\left(\beta_{\rm UV}+2\right)\,\log\,\left(\left(1+z\right){1600\,{\textrm{\AA}}}/\lambda_{\rm obs}\right)+2.5\,\log\,\left(1+z\right)$
		where DM is the distance modulus for the assumed cosmology.}
\tablenotetext{3}{Luminosity includes a correction for IGM and Ly$\alpha$ absorption.}
\tablenotetext{4}{The H$\alpha$ emission line is double peaked with centers of
		peak at	$z=1.5585$ and $z=1.5597$. Without loss of generality we
		assume $z=1.559$.}
\tablenotetext{5}{The host identification is not unique.}
\tablenotetext{6}{The redshift range of 060923A was constructed by combining
		limits from \citet{Jakobsson2012a} and \citet{Perley2013a}.}
\tablenotetext{7}{For hosts with redshift limits, we report the UV properties at
		$z=3.5$. For the hosts of GRBs 060919, 060923C, 070621, and
		070808, the luminosities are strictly speaking upper limits.
		Note, the redshift of 060923C is between $z=0.86$ and $z=3.5$.
		For GRBs 050726 ($z<5.5$), 050803 ($z<6.1$), 050922B ($z<6.1$),
		061004 ($z<10$), and 070330 ($z<5.5$) we report their
		luminosities at $z=3.5$ to avoid corrections for IGM and Ly$\alpha$
		absorption.}
\tablenotetext{8}{Listed are the two hosts for which new redshift constraints
		were obtained through SED fitting. As discussed in Section
		\ref{sec:photoz} a broad range of physical galaxy parameters and
		redshifts fits the data within their uncertainties. The likely
		redshift of GRB\,050803 is $z\sim3.5$ and of GRB\,050922B $z\sim4.5$.}
		
\label{tab:host_mag}
\end{table*}

The VLT $R$-band data build the foundation of this paper. We complement
this dataset with {\hst} observations via a dedicated GRB host program (PI:
Levan), which targeted nearly all $3<z<4$ hosts. At $z\lesssim0.9$ the TOUGH
$R$-band data probe the rest-frame optical, and for these hosts we obtained data
in bluer filters with TNG/LRS, Keck/LRIS, GTC/OSIRIS, and Gemini-S/GMOS, or used
archival data. A log of the data is shown in Table\ \ref{tab:obs_log}. In addition,
for the fields of GRBs 050803 and 050922B (two GRBs with uncertain redshifts),
we succeeded in obtaining multi-band data (Table\ \ref{tab:obs_log}) probing the
spectral energy distribution (SED) from the rest-frame UV to the near-IR. Their \spitzer\
observations are described in detail in \citet{Perley2015b, Perley2015a}.

To secure the field calibration of GRB\,060805A we used the 60-inch
Palomar telescope, and of GRB\,060729 with the Gamma-ray
Optical/Near-infrared Detector (GROND; \citealt{Greiner2008a}) mounted at the
MPG/ESO 2.2 m telescope.

Furthermore, we incorporated measurements reported in the literature; specifically
we used \citet{Perley2009a} for GRB 050416A, \citet{Chen2009a} for 050820A,
\citet{Hjorth2012a} for the $K_s$ band photometry for GRBs 050803 and 050922B,
\citet{Mangano2007a} for 060614, \citet{Kruehler2011a} for 070306, and
\citet{Tanvir2012a} for GRBs 050904, 060522, and 060927.

The ground-based data were reduced in a standard fashion (bias subtraction,
flat fielding, co-adding) with dedicated software packages (Keck: customized IDL
routine, Gemini: \texttt{Gemini IRAF package} package, GROND: customized pipeline
\citep[for details see][]{Yoldas2008a,Kruehler2008a}, and all other data with
\texttt{IRAF}; \citealt{Tody1986a}). \hst\ observations, after standard
``on-the-fly" processing, were subsequently cleaned for bias striping, introduced
due to the replacement electronics after Servicing Mission 4 (2009 May 11-24),
and then drizzled with the \texttt{multidrizzle} software \citep{Koekemoer2003a}
into final science images. The reduction of the \spitzer\ data is described in
\citet{Perley2015a}.

\section{Methods}

\begin{figure*}
\begin{center}
\includegraphics[width=0.245\textwidth]{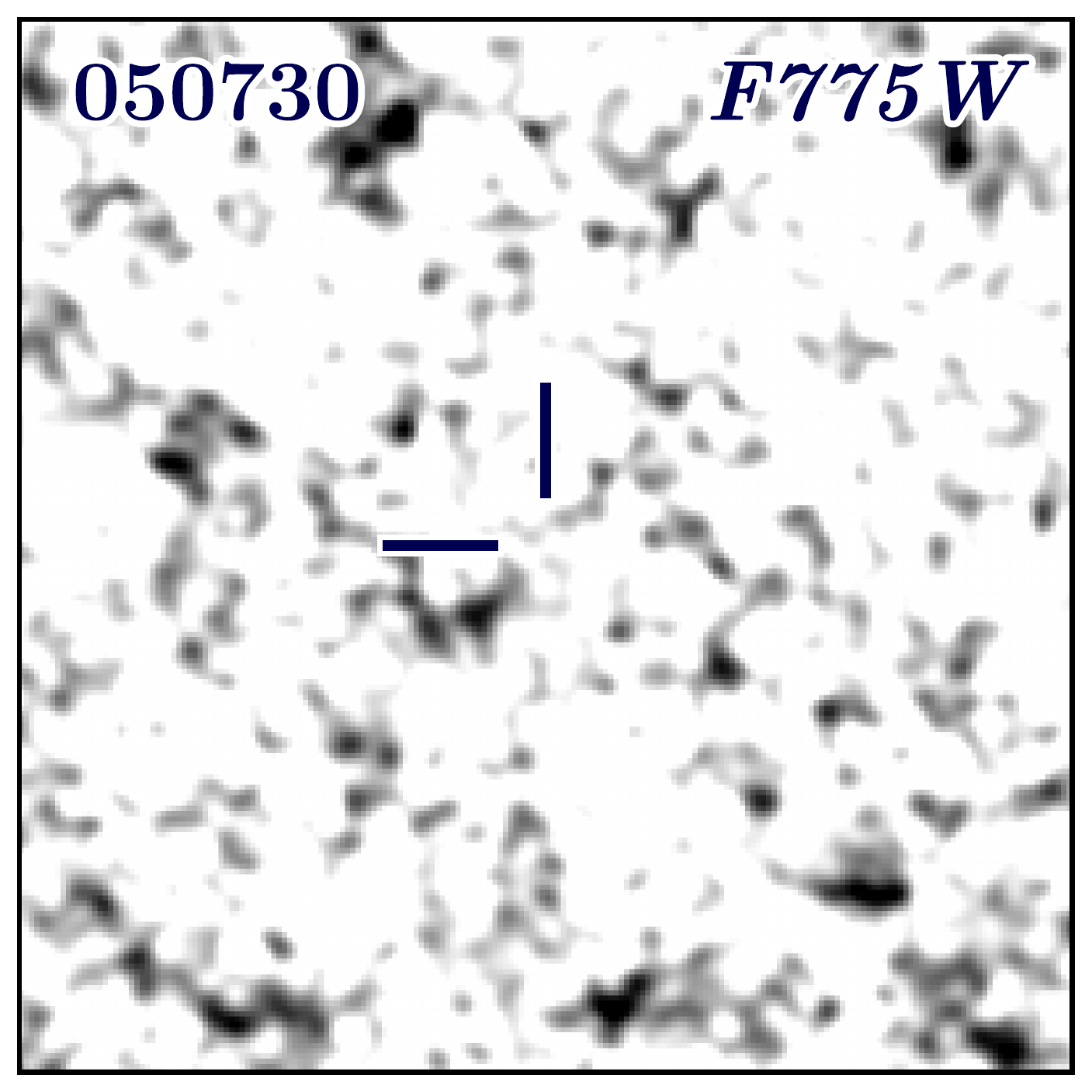}
\includegraphics[width=0.245\textwidth]{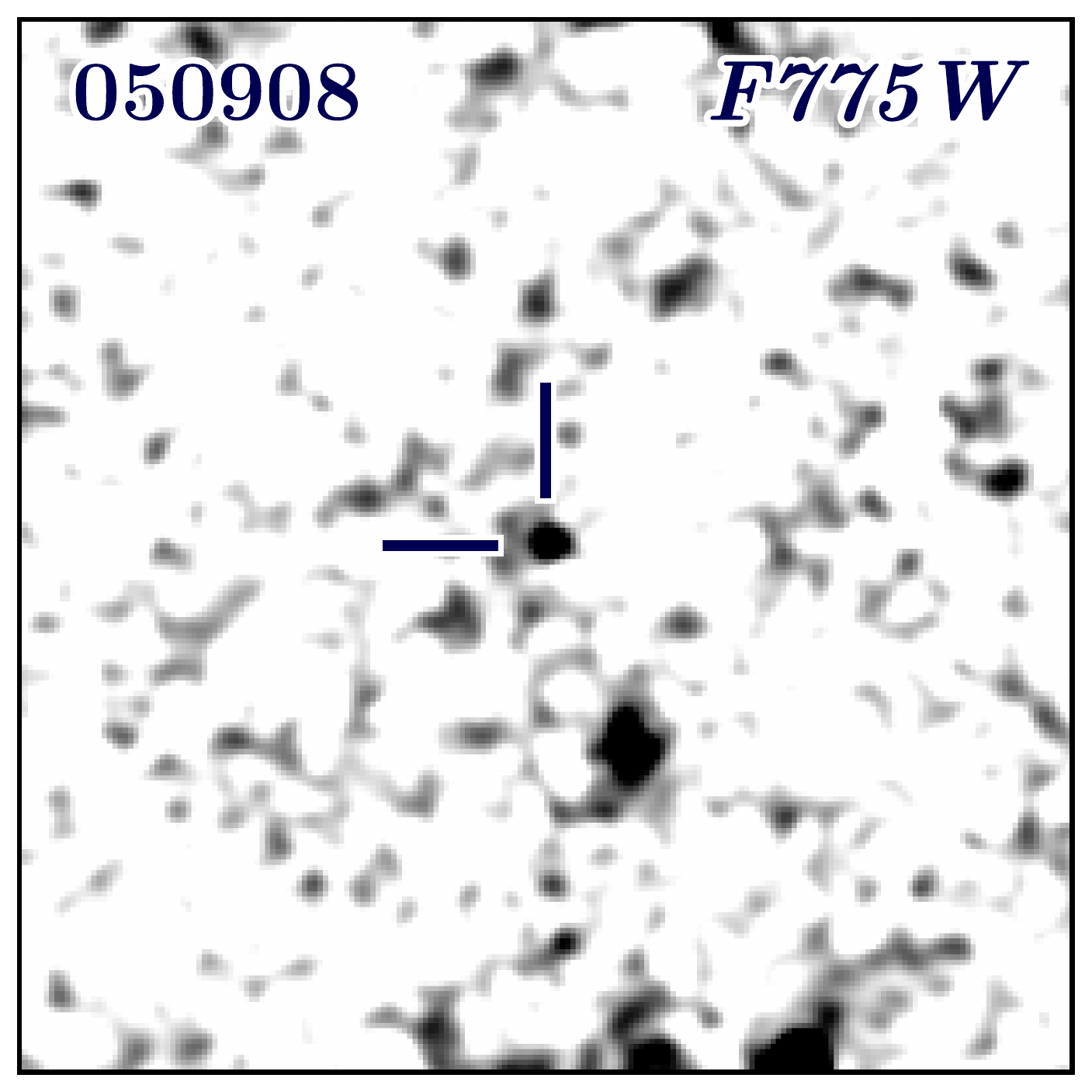}
\includegraphics[width=0.245\textwidth]{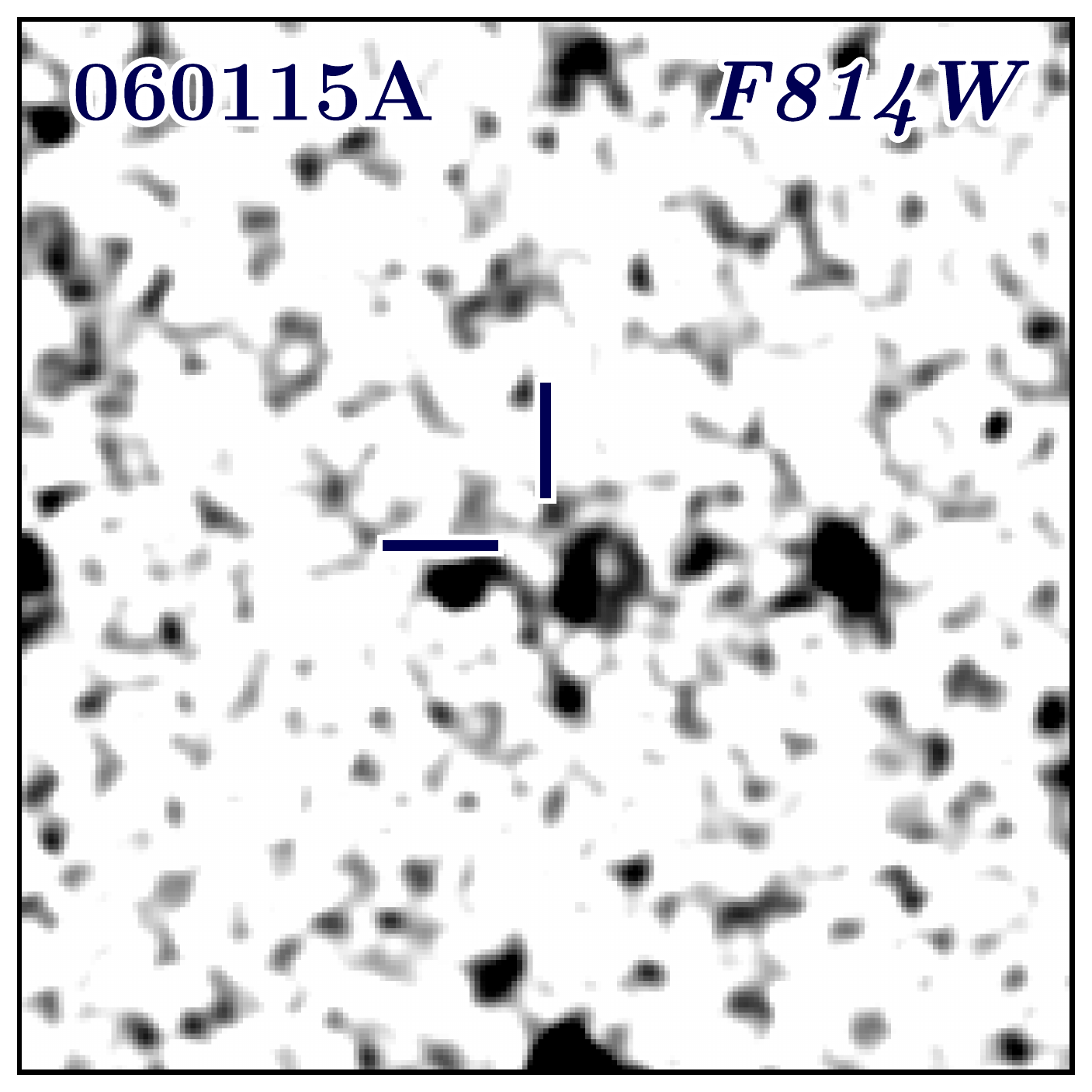}
\includegraphics[width=0.245\textwidth]{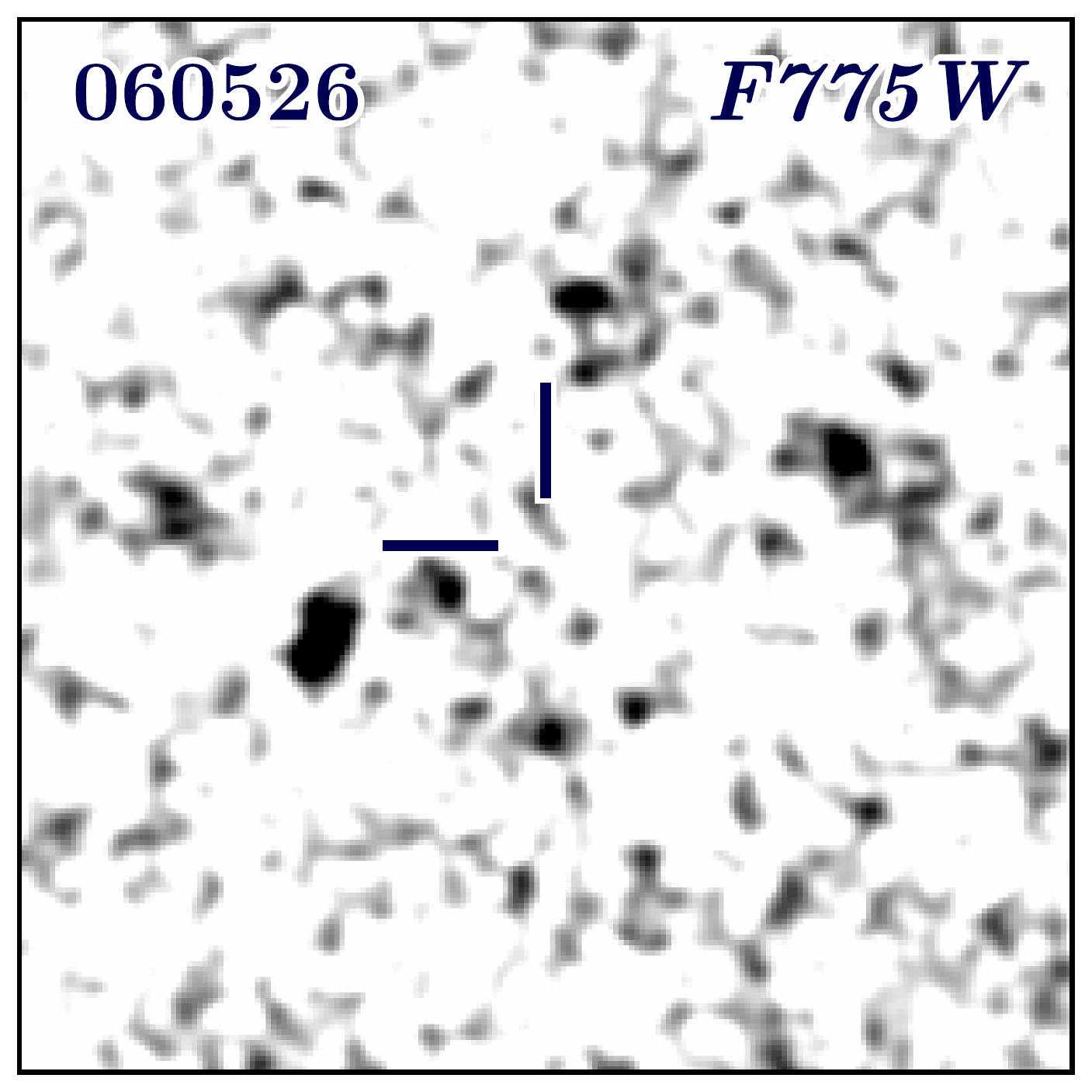}
\includegraphics[width=0.245\textwidth]{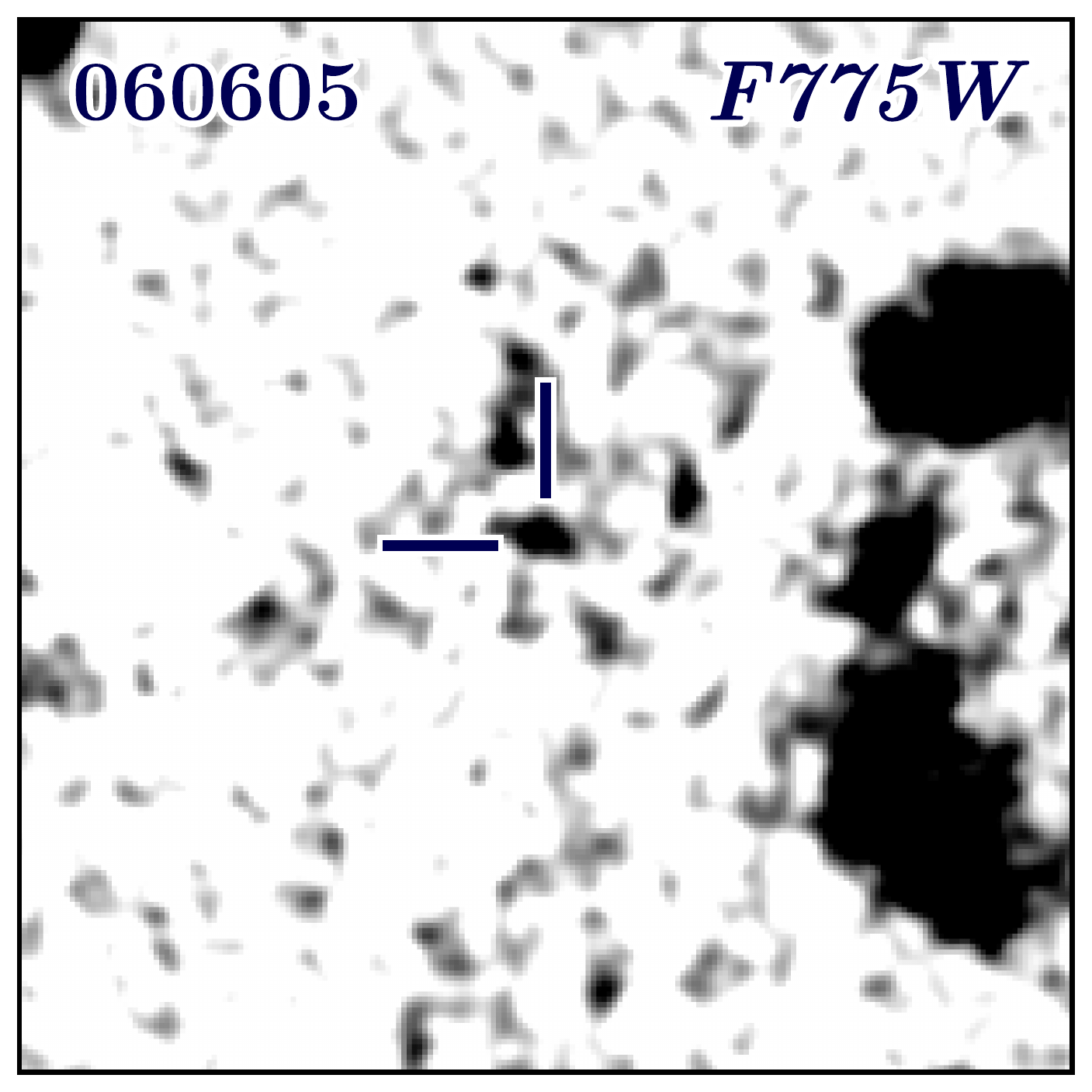}
\includegraphics[width=0.245\textwidth]{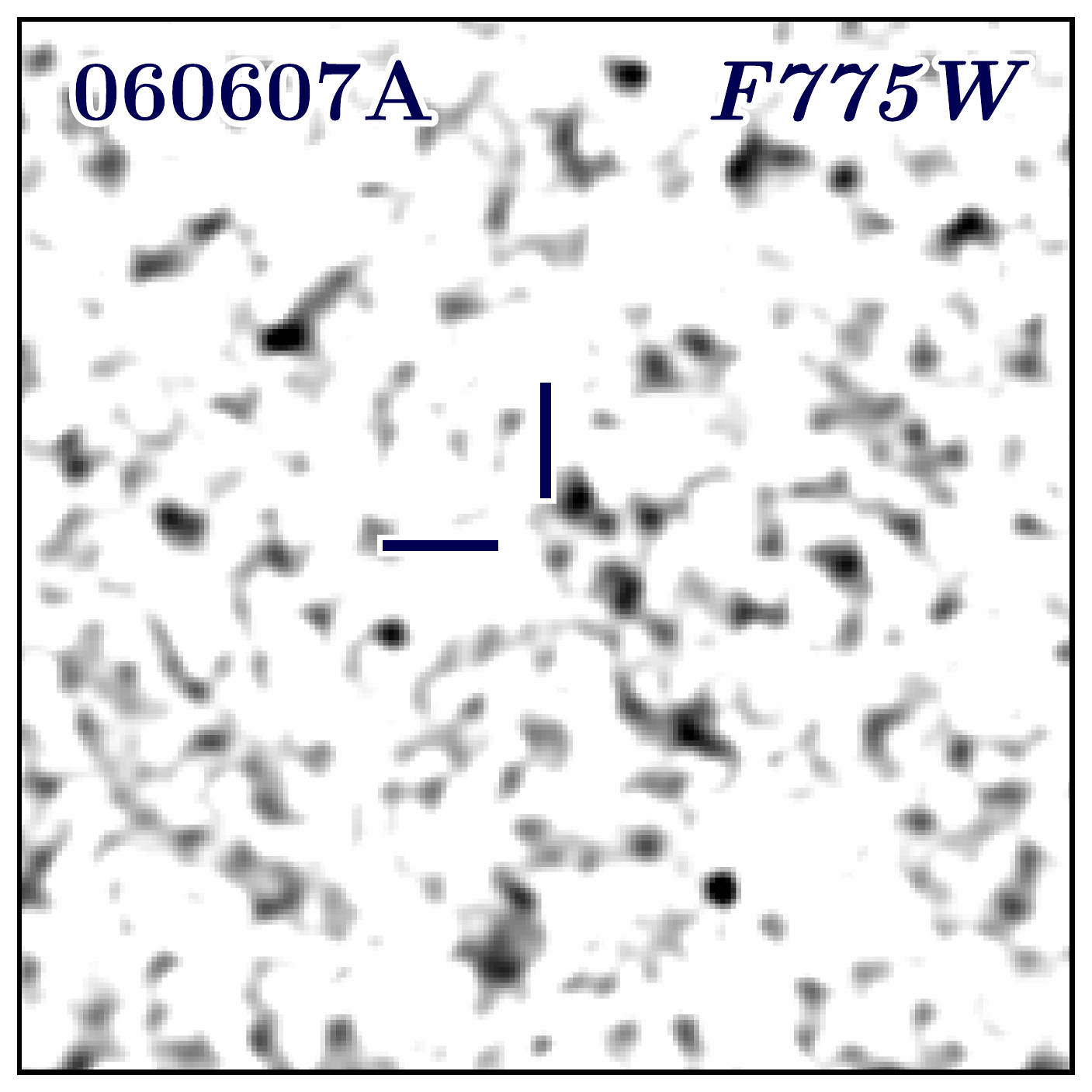}
\includegraphics[width=0.245\textwidth]{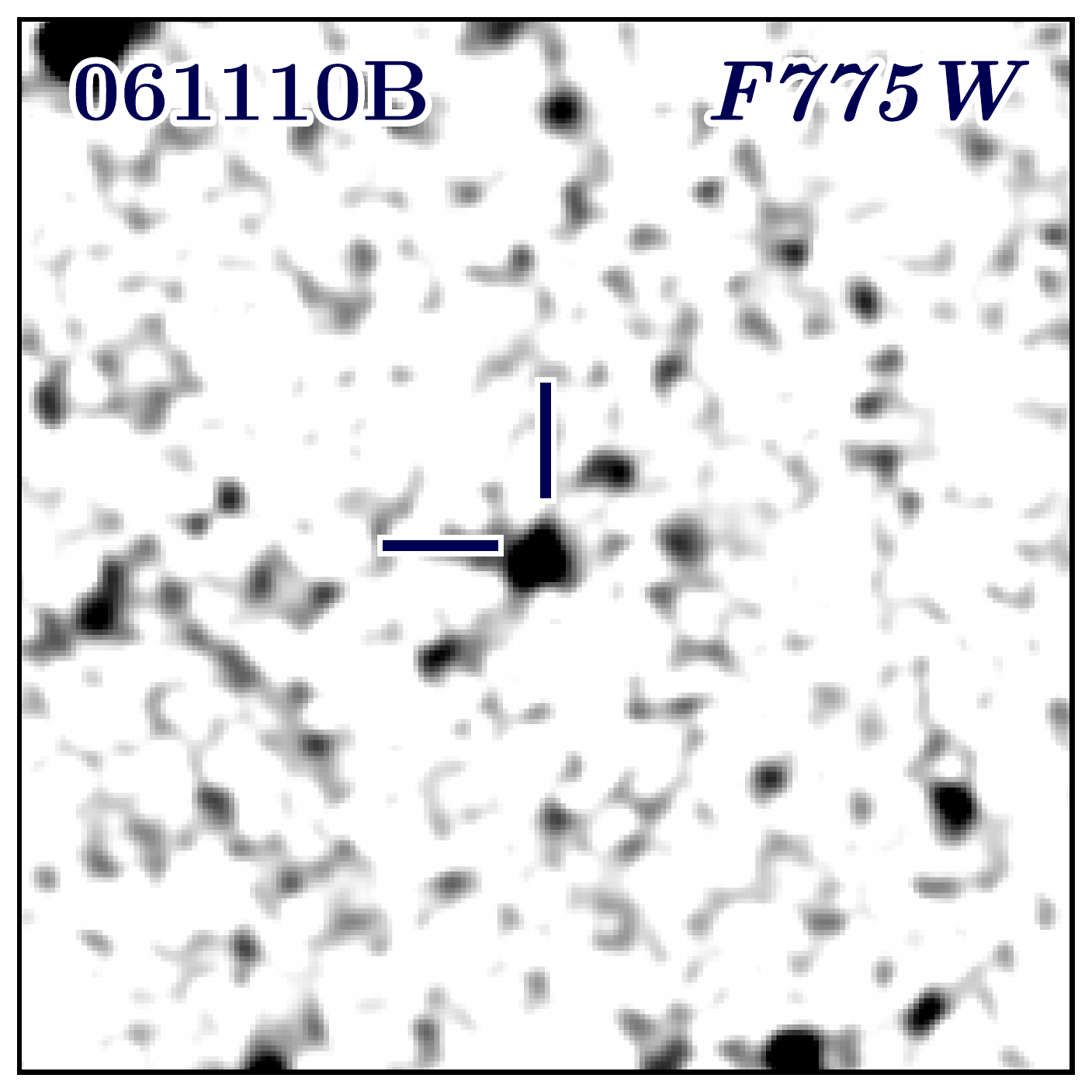}
\includegraphics[width=0.245\textwidth]{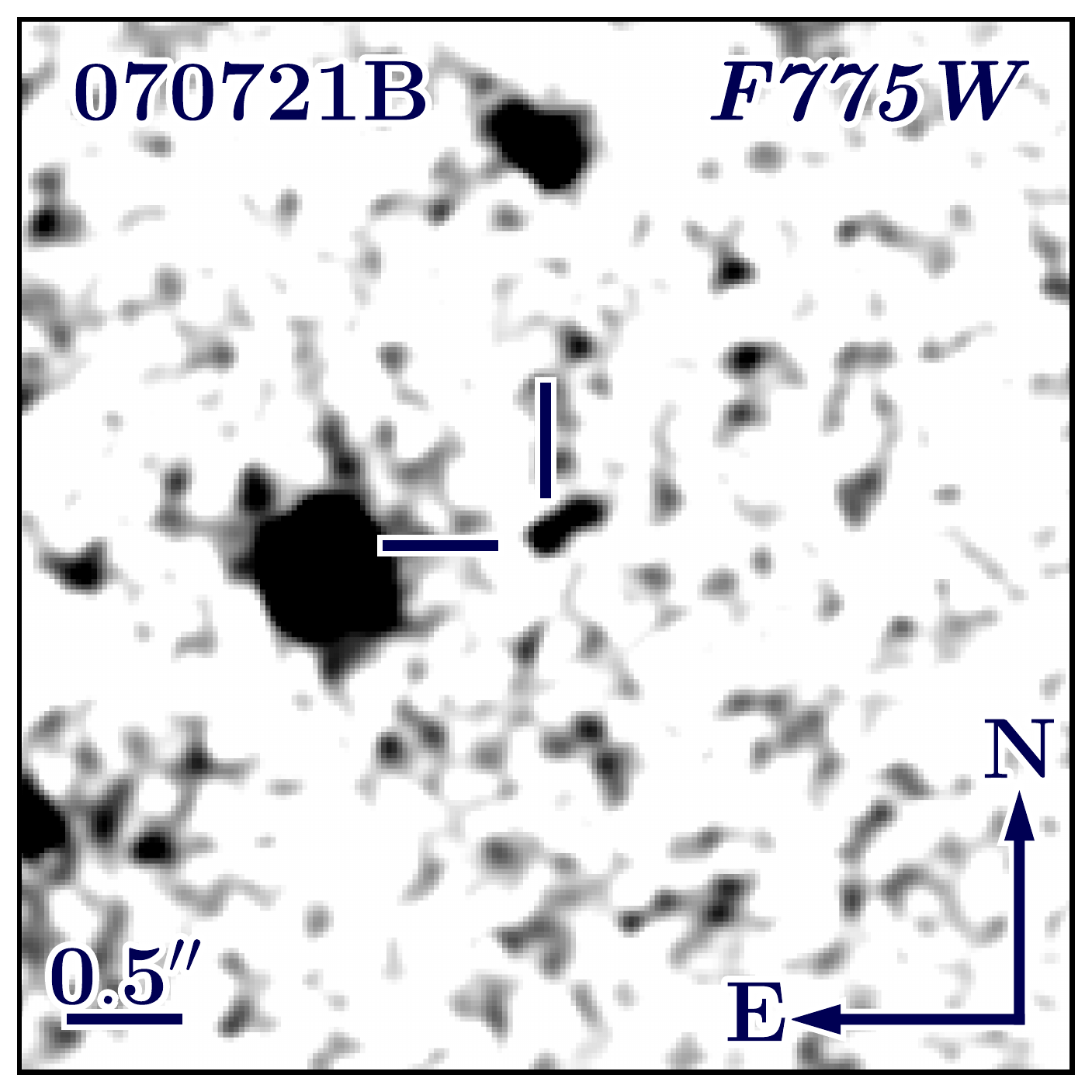}
\caption{Poststamps of the fields recently observed with {\hst}/ACS. Each cutout has
a size of $5''\times5''$, corresponding to $\sim37.3\times37.3~\rm{kpc}^2$ at $z=3.5$.
The blue crosshairs mark the positions of the optical afterglow. 
About $1''$ east of the afterglow of GRB\,070721B
is a galaxy with $m(F775W)\sim24.4$~mag ($p_{\rm ch}\sim2\%$). In
\citet{Schulze2012a}, we showed that this galaxy is in fact the galaxy counterpart
of the intervening DLA at $z=3.0939$ reported in \citet{Fynbo2009a}. For
presentation purposes all images were smoothed with a Gaussian kernel with a width
of $0.05''$.}
\label{fig:HST}
\end{center}
\end{figure*}

\subsection{Photometry}

The photometry for the ground-based images was performed as described in Malesani
et al. (2015, in preparation): if a host was detected at $>5\sigma$ confidence level, we chose
a sufficiently large aperture to measure the total flux. For hosts detected at a 
lower confidence level, we reduced the radius to the stellar FWHM and applied an
aperture correction derived from the brighter hosts in the sample.

Once an instrumental magnitude was established it was photometrically calibrated
against zeropoints (GRB\,051117B), against the brightness of a photometric
standard star (GRB\,050525A), a number of field star measured in a similar way
(GRBs\,060729), or tied to the SDSS DR8 \citep[the rest;][]{Aihara2011a}.
In some cases, we converted the SDSS photometry into the Bessel system using
Lupton (2005),\footnote{\href{http://www.sdss.org/dr5/algorithms/sdssUBVRITransform.html}{http://www.sdss.org/dr5/algorithms/sdssUBVRITransform.html}} if needed.
These measurements were finally corrected for Galactic extinction using the
extinction maps by \citet{Schlegel1998a} and transformed into the AB system
using \citet{Blanton2007a}, and \citet{Breeveld2011a}. These magnitudes are
listed in Table \ref{tab:host_mag}.

The $R$-band observation of the host galaxy of GRB 050502B is affected by
Ly$\alpha$ absorption in the host galaxy and in the intergalactic medium (IGM).
To quantify this attenuation, we compared the observed to the expected
$\left(R-I\right)$ color of the afterglow. \citet{Afonso2011a} reported that
the X-ray-to-optical SED of the afterglow can be
described by a simple power law, $F_\nu\propto\nu^{-\beta}$, with a spectral
index of $\beta\sim0.9$. For this model, the expected $\left(R-I\right)_{\rm AB}$
color is 0.22~mag, i.e., the emission received in the $R$ band is dimmed by
0.95~mag, assuming no reddening at the GRB site. Throughout the paper we assume
that the host galaxy has the same attenuation. The values reported in Table
\ref{tab:host_mag} include this correction.

For the \hst\ images we measured the background-subtracted flux within an aperture
with a radius of 0\farcs25. To quantify the measurement error, we randomly
distributed 40 apertures of the same radius within $3''$ from the optical afterglow.
These apertures had a minimum distance of 0\farcs6 from any object to avoid source
contamination. After that, we computed the standard deviation as a proxy for the
photometric error. To account for flux loses we applied aperture corrections that
were calculated from the encircled energy \citep{Sirianni2005a}. If no host was
detected, we measured the $3\sigma$ limiting magnitude via
$m_{\rm limit}=23.9-{\rm{AP}}_{\rm cor}-2.5\,\log\left(F_\nu + 3\,\sigma\left(F_\nu\right)\right)$
where $\rm{AP}_{\rm cor}$ is the aperture correction and $F_\nu$ is the formal
flux density in $\mu$Jy measured at the position of the optical afterglow with
its $1\sigma$ error $\sigma\left(F_\nu\right)$.

At $z\sim3$, an angular size of 0\farcs5 translates to physical scale of $\sim4$~kpc.
This diameter is at the lower end of the observed size distribution of LBGs
\citep{Hathi2008a}. Increasing the aperture radius to 0\farcs3, the average size of a
LBG at $z\sim3$, leads to no significant increase in flux. On the contrary,
the measurement would have been affected by neighboring objects if the radius
exceeded 0\farcs4.

The analysis of the \spitzer\ data is described in detail in \citet{Perley2015a};
in brief, after downloading the processed PBCD images from the
Spitzer Legacy Archive,\footnote{\href{http://sha.ipac.caltech.edu/applications/Spitzer/SHA/}{http://sha.ipac.caltech.edu/applications/Spitzer/SHA/}}
we modeled and subtracted nearby contaminating sources, and then measured the
flux of the host via aperture photometry and used the IRAC handbook
zeropoints to convert the instrumental to apparent magnitudes.\footnote{\href{http://irsa.ipac.caltech.edu/data/SPITZER/docs/irac/iracinstrumenthandbook/}{http://irsa.ipac.caltech.edu/data/SPITZER/docs/irac/\\ iracinstrumenthandbook/}}

\subsection{Host Identification}\label{sec:host_id}

\begin{table}
\caption{Properties of the $z=3$--4 Host Galaxies Observed With \hst}
\centering
\scriptsize
\begin{tabular}{c@{\hspace{1mm}}c@{\hspace{1mm}}c@{\hspace{1mm}}c@{\hspace{1mm}}c@{\hspace{1mm}}c@{\hspace{1mm}}c@{\hspace{1mm}}c@{\hspace{1mm}}c}
\toprule
\multirow{2}{*}{GRB}	& $\alpha,~\delta$& $r_{1/2}$		& $r_0$			& \multirow{2}{*}{Band}		& $F_\nu$	 	& Brightness 				& \multirow{2}{*}{$p_{\rm ch}$}	\\
			& (J2000)	& $('')$		& $('')$		& 				& $\left(\rm{nJy}\right)$& (mag)				&				\\
\midrule	
		050730	& \nodata	& \nodata		& \nodata		& $F775W$			& $7\pm7$			& $>27.6$				&\nodata			\\[2mm]
\multirow{2}{*}{050908}	& 01:21:50.727,	& \multirow{2}{*}{0.05}	& \multirow{2}{*}{0.02}	& \multirow{2}{*}{$F775W$}	& \multirow{2}{*}{$29\pm7$}	& \multirow{2}{*}{$27.60^{+0.30}_{-0.24}$}& \multirow{2}{*}{0.002}	\\
			& -12:57:17.31	&			&			&				& 				&					&				\\[2mm]
\multirow{4}{*}{060115}	& 03:36:08.314,	& \multirow{2}{*}{0.09}	& \multirow{2}{*}{0.28}	& \multirow{2}{*}{$F814W$}	& \multirow{2}{*}{$38\pm7$}	& \multirow{2}{*}{$27.27^{+0.22}_{-0.18}$}& \multirow{2}{*}{0.020}	\\
			& +17:20:42.80	&			&			&				& 				&					&				\\[1mm]
			& 03:36:08.351,	& \multirow{2}{*}{0.08}	& \multirow{2}{*}{0.44}	& \multirow{2}{*}{$F814W$}	& \multirow{2}{*}{$26\pm7$}	& \multirow{2}{*}{$27.68^{+0.34}_{-0.26}$}& \multirow{2}{*}{0.052}	\\
			& +17:20:42.86	&			&			&				& 				&					&				\\[2mm]
		060526	& \nodata	& \nodata		& \nodata		& $F775W$			& $5\pm7$			& $>27.66$				& \nodata			\\[2mm]
\multirow{4}{*}{060605}	& 21:28:37.312,	& \multirow{2}{*}{0.09}	& \multirow{2}{*}{0.06}	& \multirow{2}{*}{$F775W$}	& \multirow{2}{*}{$30\pm9$}	& \multirow{2}{*}{$27.55^{+0.38}_{-0.28}$}& \multirow{2}{*}{0.007}	\\
			& -06:03:30.96	&			&			&				& 				&					&				\\[1mm]
			& 21:28:37.321,	& \multirow{2}{*}{0.06}	& \multirow{2}{*}{0.47}	& \multirow{2}{*}{$F775W$}	& \multirow{2}{*}{$28\pm9$}	& \multirow{2}{*}{$27.63^{+0.42}_{-0.30}$}& \multirow{2}{*}{0.054}	\\
			& -06:03:30.56	&			&			&				& 				&					&				\\[2mm]
		060607$^{\rm a}$	& \nodata	& \nodata		& \nodata		& $F775W$			& $13\pm8$			& $>27.48$				& \nodata			\\[2mm]
\multirow{2}{*}{061110B}& 21:35:40.396,	& \multirow{2}{*}{0.12}	& \multirow{2}{*}{0.05}	& \multirow{2}{*}{$F775W$}	& \multirow{2}{*}{$45\pm10$}	& \multirow{2}{*}{$27.10^{+0.26}_{-0.21}$}& \multirow{2}{*}{0.008}	\\
			& +06:52:34.30	&			&			&				& 				&					&				\\[2mm]
\multirow{2}{*}{070721B}& 02:12:32.935,	& \multirow{2}{*}{0.08}	& \multirow{2}{*}{0.20}	& \multirow{2}{*}{$F775W$}	& \multirow{2}{*}{$25\pm9$}	& \multirow{2}{*}{$27.76^{+0.47}_{-0.33}$}& \multirow{2}{*}{0.018}	\\
			& -02:11:40.63	&			&			&				& 				&					&				\\
\bottomrule
\end{tabular}
\tablecomments{
	For each galaxy we list its half-light radius $r_{1/2}$, its projected
	distance to the GRB $r_0$, its flux density $F_\nu$, and the apparent
	magnitude. If no host candidate was detected, we report the nominal
	flux density at the afterglow position and the corresponding $3\sigma$
	limiting magnitude. All magnitudes (but not the flux densities) include
	an aperture correction but \textit{no} correction for Galactic reddening.
	The uncertainty in the reported coordinates is $\sim0\farcs4$ (comprising
	the astrometry error of the optical afterglow images and the alignment
	error of the VLT and \hst\ images). See Section \ref{sec:host_id} for details.
}
\tablenotetext{1}{
	There is in fact a host candidate with a chance probability of
	$p_{\rm ch}=0.04$ 0\farcs29 from the afterglow. However it is only detected
	in a very small aperture with a radius of 0\farcs2. The coordinates of
	the object are R.A., decl.(J2000)= 21:58:50.388, -22:29:46.68 $\pm$
	0\farcs4. Its magnitude is $m(F775W)=28.28^{+0.53} _{-0.35}$~mag.
}
\label{tab:host_cand}
\end{table}

Malesani et al. (2015, in preparation) describes in detail how the hosts were identified in
the deep VLT images. The additional data obtained with ground-based telescopes
have a similar spatial resolution and do not exceed the limiting magnitudes of
the VLT images. In contrast the \hst\ images exceed the VLT images in spatial
resolution and depth (see Figure \ref{fig:HST}). This necessitates repeating the
host identification. In the following we can limit the discussion to the $z=3$--4
hosts, whereas  the host identification of GRB 050820A is discussed in
\citet{Chen2009a} and \citet{Chen2012a}, and of GRBs 050904, 060522, and
060927 in \citet{Tanvir2012a}.

To identify the most likely host galaxy candidate, we chose the probabilistic
approach by \citet{Bloom2002a} \citep[For a detailed discussion see also][]{Perley2009a}.
This method is based on the observed galaxy density from \citet{Hogg1997a} and
quantifies the chance probability $p_{\rm ch}$ of finding a galaxy with a certain
magnitude and a certain distance from the GRBs. This chance probability is given
by
\begin{equation}
p_{\rm ch} = 1-\exp\left(-\pi\times r_{\rm eff}\times\sigma\left(m\right)\right)\nonumber
\end{equation}
where $r_{\rm eff}$ is the effective radius and $\sigma\left(m\right)$ is the
galaxy density for a given observed magnitude $m$. The effective radius depends
on the  half-light radius $r_{1/2}$, the distance from the GRB $r_0$, and
localization accuracy. The localization accuracy is defined by the error of
aligning the \textit{Hubble Space Telescope} (\hst) images to images of the optical afterglows which is between
0\farcs033--0\farcs061. The effective radius in \citet{Bloom2002a} can hence be
re-written as $r_{\rm eff}=2\,r_{1/2}$ if $r_0<r_{1/2}$ and
$r_{\rm eff}=(r^2 _0 + 4\,r^2 _{1/2})^{1/2}$ if $r_0>r_{1/2}$. The half-light
radii were measured with \texttt{SExtractor} v2.19.5 \citep{Bertin1996a}. To
limit the number of candidates we set an upper limit of $3''$ on the GRB offset
and required a chance probability of $p_{\rm ch}<5\%$.

The final candidates are listed in Table \ref{tab:host_cand} (not corrected for
Galactic reddening, whereas the unreddended magnitudes are reported in Table
\ref{tab:host_mag}). The host offsets are consistent with the observed
distribution for the TOUGH sample (Malesani et al. 2015, in preparation). If no host
candidate was detected, we report the nominal flux and the $3\sigma$ limiting
magnitude. The host identifications of GRBs 060115 and 060605 are not unique,
where alternative candidates for each of these GRBs have the same magnitudes
within $2\sigma$. For simplicity, we used their weighted means in the further
analysis. 

\subsection{Photometric Redshifts}\label{sec:photoz}

Although the TOUGH sample has a current redshift completeness of 87\%, nine
hosts remain without precise redshift information (Table\,\ref{tab:host_mag}).
We succeeded in obtaining multi-band data for GRBs 050803 and 050922B from
4000 to 42000~\AA\ (Table\,\ref{tab:obs_log}) to model their SEDs and obtain
photometric redshifts.

The field of GRB 050803 was observed in the same filters but with a different
telescope (Table\,\ref{tab:host_mag}). We built super-stacks for each band by
resampling these data to the grid of the Keck/LRIS images (which has the highest
spatial resolution) while conserving the flux and weighting the images by their
limiting magnitudes. The final extinction-corrected magnitudes of the two hosts
are reported in Table \ref{tab:photoz}.

\begin{table}
\caption{Broad-band Photometry of GRBs 050803 and 050922B}
\centering
\scriptsize
\begin{tabular}{cc@{\hspace{1cm}}cc}
\toprule
\multicolumn{2}{c}{\textbf{GRB 050803}}		& \multicolumn{2}{c}{\textbf{GRB 050922B}}\\
\midrule
\multicolumn{1}{l}{\multirow{2}*{Band}}	& Brightness	& \multicolumn{1}{c}{\multirow{2}*{Band}}	& Brightness\\
							& (mag)			&			& (mag)\\
\midrule
$g'$			& $>27.45$			& $g'$			& $27.50\pm0.50$\\
$R$			& $26.29 \pm 0.22$		& $R$			& $26.52\pm0.22$\\
$i'$			& $26.45 \pm 0.50$		& $i'$			& $25.18\pm0.14$\\
$F160W$			& $25.74 \pm 0.18$		& $z'$			& $25.01\pm0.34$\\
$K_s$			& $>23.30$			& $K_s$			& $>24.00$	\\
$3.6~\mu\rm m$		& $>24.67$			& $3.6~\mu\rm m$	& $24.77\pm0.36$\\
$4.5~\mu\rm m$		& $>25.00$			& $4.5~\mu\rm m$	& $24.60\pm0.46$\\
\bottomrule
\end{tabular}
\tablecomments{All magnitudes are corrected for
Galactic reddening. Non-detections are reported at $3\sigma$ confidence level.
The $K$-band photometry was taken from \citet{Hjorth2012a}.}
\label{tab:photoz}
\end{table}

\begin{figure}
\begin{center}
\includegraphics[width=1.0\columnwidth]{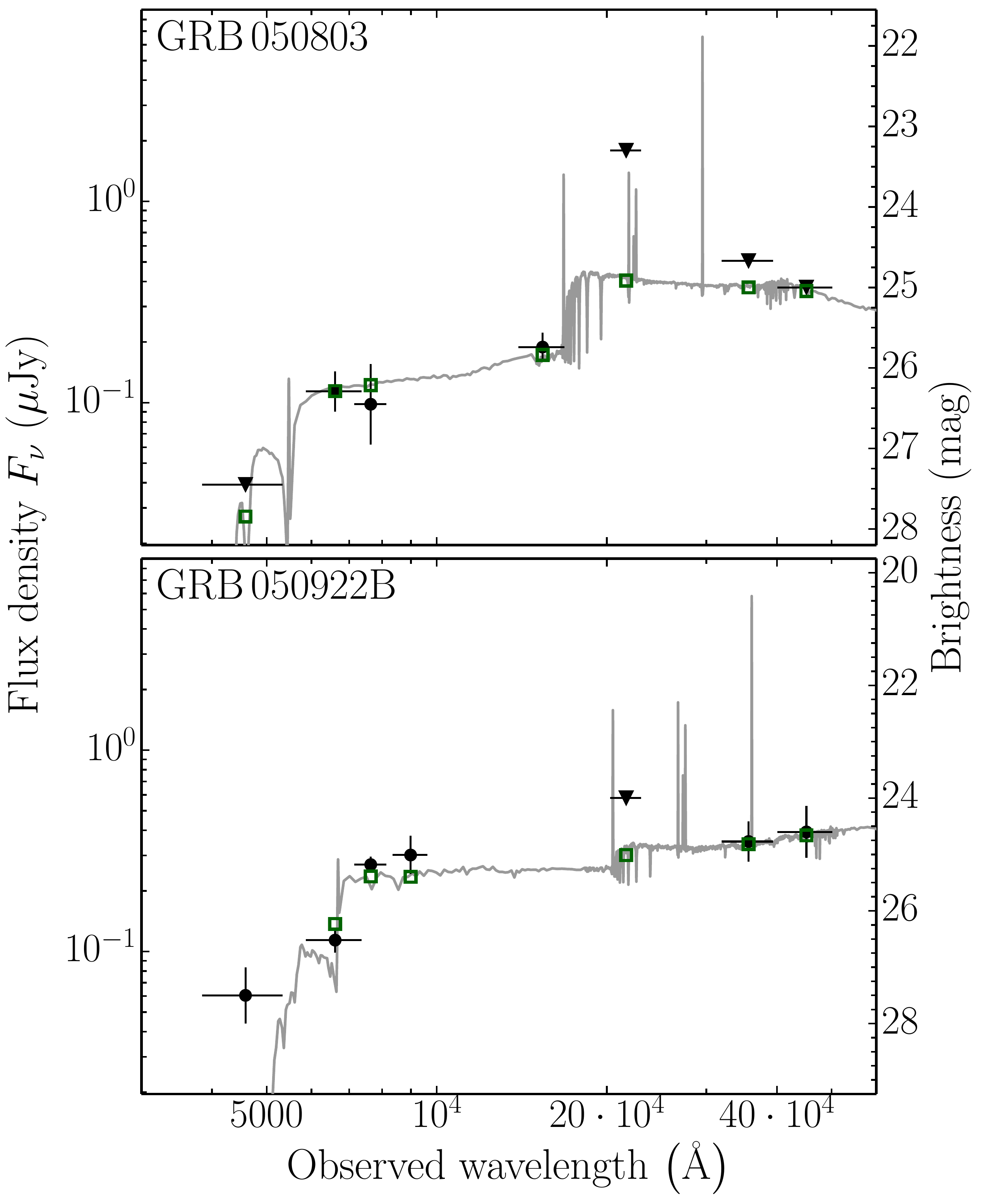}
\caption{Spectral energy distributions (SEDs) of GRBs 050803 and 050922B
and their fits. The solid line displays a fit of the SED with Le Phare. The green
open squares are the model predicted magnitudes. Given the sampling and the 
measurement errors, a broad range of physical galaxy parameters and redshifts fits
the data within their uncertainties. Assuming the two galaxies to be star-forming
the most likely redshift of GRB 050803 is between $z=3$ and 4 and of GRB 050922B
between $z=4$ and 5. See Section \ref{sec:photoz} for details.}
\label{fig:photoz}
\end{center}
\end{figure}

We modeled the SED with \texttt{Le Phare} \citep{Arnouts1999a, Ilbert2006a},\footnote{\href{http://www.cfht.hawaii.edu/~arnouts/LEPHARE}{http://www.cfht.hawaii.edu/\~{}arnouts/LEPHARE}}
using a grid of galaxy templates based on \citet{Bruzual2003a} stellar 
population-synthesis models with the Chabrier IMF and a Calzetti dust attenuation
curve \citep{Calzetti2000a}. For a description of the galaxy templates, physical
parameters of the galaxy fitting and their error estimation, we refer to
\citet{Kruehler2011a}. To account for zeropoint offsets in the cross calibration
and absolute flux scale, a systematic error contribution of 0.05 mag was added
in quadrature to the uncertainty introduced by photon noise. Figure\,\ref{fig:photoz}
displays the observed host SEDs and their best fits.

Considering the brightness of the two galaxies ($R>26\,\rm{mag}$) and their
measurement errors, this fitting does not yield a unique redshift solution for
either of the two galaxies. A broad range of physical galaxy parameters and
redshifts fit the data within their uncertainties. Of particular interest in our
case is the possibility that the galaxies are at $z > 3$. A high redshift does
in fact provide a good description of the photometry in both cases. The SED of
the host of GRB~050922B has a significant $g'-r'$ and $r'-i'$ color which are
reasonably-well explained by the Ly-$\alpha$ and Lyman-limit breaks at $z\sim4.5$.
Similarly, the red $g'-r'$ and blue $r'-i'$ color of the galaxy hosting
GRB~050803 is indicative of a redshift of $z\sim3.5$. Lower redshift solutions
($z<1$) exist as well in both cases, but based on the available photometry, it
is at least plausible that both GRBs originated at $z>3$.

\begin{figure}
\begin{center}
\includegraphics[width=1.0\columnwidth]{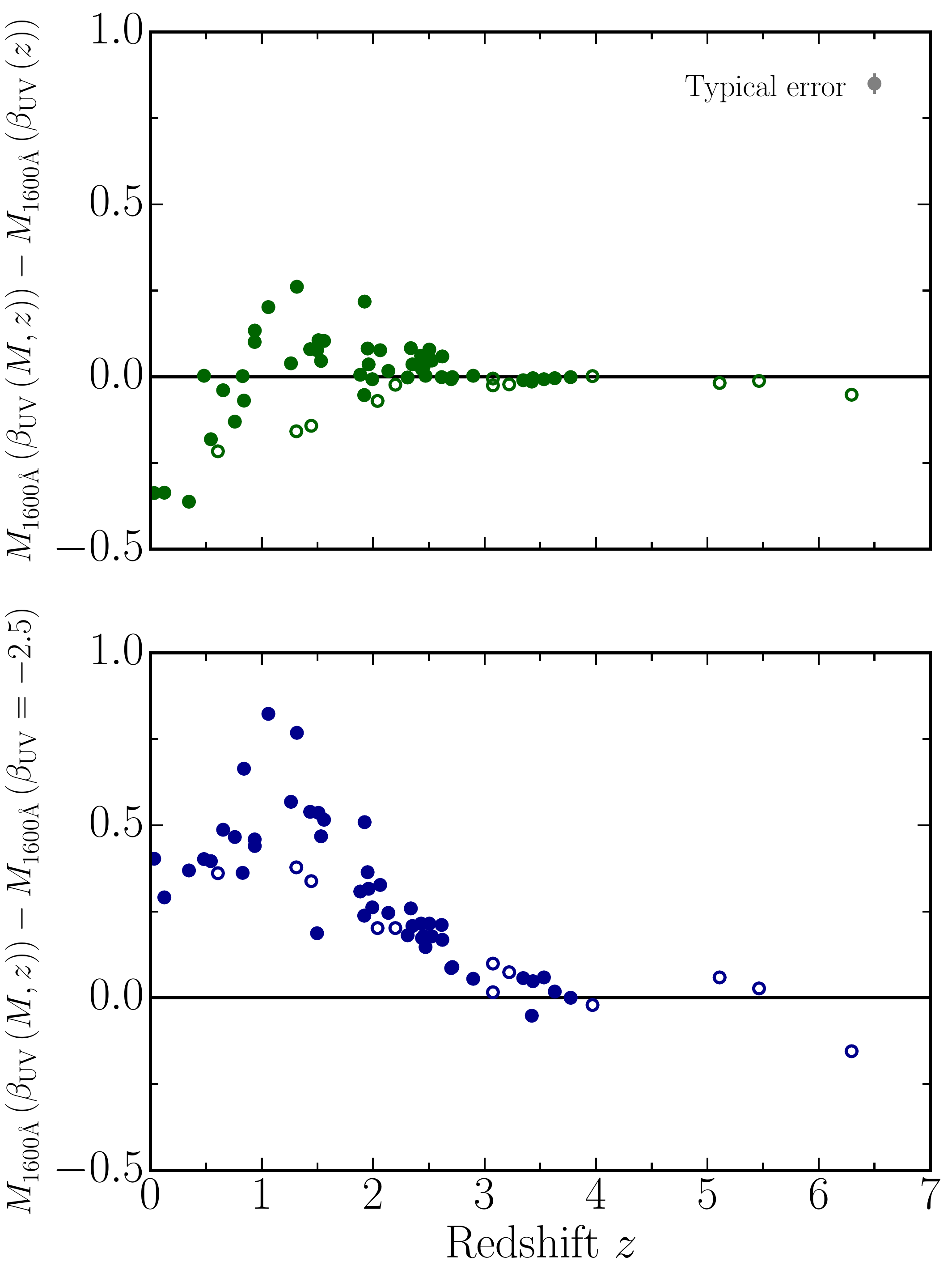}
\caption{Impact of the approximations on the rest-frame UV continuum on the
	observed UV luminosity at 1600~\AA. The continuum is assumed to be a
	power-law shaped with $F_\lambda\propto\lambda^{-\beta_{\rm UV}}$,
	where the slope can be luminosity and redshift dependent.
	\textit{Top panel}: luminosity- and redshift-dependent slope vs.
	luminosity-independent but redshift-dependent slope. \textit{Bottom
	panel}: luminosity- and redshift-dependent slope vs. uniform slope.
	Detected hosts are displayed as filled circles and non-detected hosts
	as open circles. For clarity, we only show the hosts with known
	redshifts. We omit the host of GRB\,050502B because of its uncertain
	IGM correction. The size of the average error is shown in the top panel.}
\label{fig:slope_comp}
\end{center}
\end{figure}

\begin{figure*}[t!]
\begin{center}
\includegraphics[width=1\textwidth]{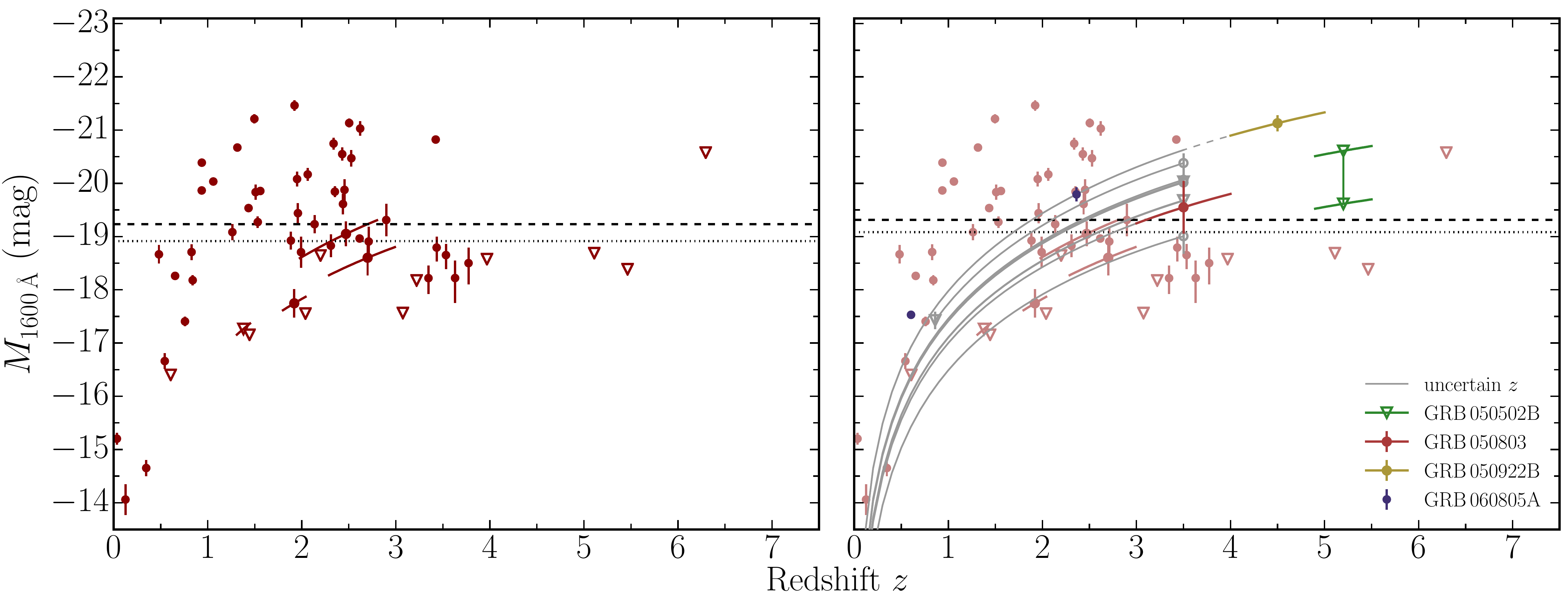}
\caption{
	The absolute magnitudes of our GRB host galaxy sample plotted vs.
	redshift. \textit{Left}: host with known redshifts of detected ($\bullet$)
	and non-detected hosts ($\triangledown$). The horizontal dashed line 
	indicates the median luminosity of detected hosts (being $-19.2$~mag)
	and the dotted line of all including the non-detected hosts (the median
	value being $-18.9$~mag). \textit{Right}: location of the hosts with
	uncertain redshifts (curved lines), GRB 060805A, which does not
	have a unique host identification, and GRB 050502B, which has an
	uncertain IGM correction. To illustrate the impact of the IGM correction,
	the parameter space ranging from no IGM correction to the 
	conservative IGM correction discussed in Section \ref{sec:data_red} is
	shown. As in the left panel, the dashed line indicates the median
	luminosity of detected hosts (being $-19.3$~mag) and the dotted line of
	all hosts (being $-19.1$~mag), where we assumed hosts without redshift
	information to be at $z=3.5$ according to Table \ref{tab:host_mag}.
	For GRBs\,050803 and 050922B, the likely redshift ranges from
	the SED modeling are highlighted. (see Section \ref{sec:photoz} for
	details). Because of the ambiguity of the redshift ranges, we also show
	their tracks if they are at lower redshifts.
}
\label{fig:scatter_1}
\end{center}
\end{figure*}

\subsection{The Obscured UV Luminosity}\label{sec:k-cor}

Although we have at least one measurement of the rest-frame UV continuum between
1216 and 4000~\AA\ for each host galaxy in our sample, these data probe different
parts of the UV continuum; on average they probe the rest-frame at 2140~\AA. UV
LFs of LBG samples are typically reported at 1500--1700~\AA\
\citep[e.g.,][]{Arnouts2005a, Reddy2009a, Oesch2010a, Bouwens2014b}. To shift
the UV luminosities of the TOUGH sample to the common rest-frame at 1600~\AA,
a $K$-correction was applied assuming the UV continuum to be power-law shaped
($F_\lambda\propto\lambda^{-\beta_{\rm UV}}$). However, LBG surveys have also
shown that the spectral slope is luminosity and redshift dependent
\citep[e.g.,][]{Bouwens2009a,Bouwens2010b}. To account for that we make use of
the parametrization by \citet{Trenti2014a} that is based on the findings for
LBGs by \citet[][and references therein]{Bouwens2012a}. Since the slope depends
on the unknown observed UV luminosity, we solve the inverse problem: we compute
the expected apparent magnitudes for a range of UV luminosities
($-30~{\rm mag}<M<-8~{\rm mag}$) at the redshift of each GRB and select the
luminosity that minimizes the difference between the observed and the expected
apparent magnitude. In Table \ref{tab:host_mag} we summarize the best-fit slopes
and luminosities.

We note that the UV luminosities in the FUV are highly sensitive to any
reddening correction. In LBG surveys it is a common practice to build the LF from
the obscured UV luminosities, and therefore without any loss of generality or
limitation in the comparison with LBG surveys, we omit any reddening correction.

\subsection{The Impact of the UV Slope on the $K$-correction}\label{sec:disc_k_cor}

The exact shape of the unobscured UV continuum is determined by the age
of the young stellar population and metallicity. \citet[][and its update in the
GHostS database]{Savaglio2009a} and \citet{Perley2013a} showed that the observed
age distribution extends from a few tens of megayears to two billion years.
Though this result is based on heterogeneous samples, the maximum age is consistent
with limits we extracted from the TOUGH sample (Malesani et al. 2015, in preparation). To
check whether the slopes used in our analysis are consistent with these ages,
we fit the UV continuum of single-age stellar population models from
\citet{Bruzual2003a} between 1350~\AA\ and 3600~\AA\ with power law models.
We find that the slopes vary between $\beta_{\rm UV}\sim-2.8$ and $\sim3.5$ for
ages between 0.005 Gyr and 2.5 Gyr, in agreement with the values derived in
Section \ref{sec:k-cor}.

We next assess how the assumption on the luminosity and redshift dependence
affects our results. We first assume the slope to be luminosity independent.
We constrain the redshift evolution using results from \citet{Schiminovich2005a},
\citet{Finkelstein2012a}, and \citet{Hathi2013a}. Between $z=0$ and $z=8$, the
UV slope could be parametrized as
\begin{equation}
\beta_{\rm UV}(z)=\left(-1.62\pm0.04\right)+\left(-0.07\pm0.01\right)\times z
\label{eq:k_corr}
\end{equation}
for luminosities between $\sim0.1~L_\star$ and $\sim1.5~L_\star$. Figure \ref{fig:slope_comp}
(top panel) displays the difference in the observed luminosity for the different
parameterizations of the UV slope. The median difference is 0.01mag and in
most cases negligible. The largest differences of 0.26mag are observed at $z\sim1$
where the $R$-band data have the largest distance from the common rest frame at
1600~\AA.

Next we drop the redshift dependency and assume a very young stellar population
with characteristic spectral slope of $\beta_{\rm UV}=-2.5$ for all hosts. The
luminosity will increase for all hosts (Figure \ref{fig:slope_comp}, bottom panel).
The largest shift of 0.82~mag is observed at $z\sim1$. However, it is very unlikely
that the majority of GRB hosts have such blue UV continua. Several low-redshift
hosts are known to have evolved young stellar populations, e.g., GRB 970228
($z=0.695$): $\textrm{age}=1.7$~Gyr; GRB 990712 ($z=0.433$):
$\textrm{age}=1.1$~Gyr; and GRB 011121 ($z=0.362$): $\textrm{age}=2.3$~Gyr
\citep{Savaglio2009a}. In conclusion, though individual hosts may be more or
less luminous than inferred from our \textit{ansatz}, we do not consider our UV
slope assumptions to have any systematic effect on the ensemble above $z\sim1$.

\section{Results}\label{sec:results}
\subsection{GRB Host Luminosity Distribution}\label{sec:results_lum}

Figure\,\ref{fig:scatter_1} (left panel) plots the absolute magnitudes at 1600~\AA\
(detections and upper limits) of the sample GRB hosts versus redshift. The absolute
magnitudes of the detected hosts span a wide range in magnitude from $-14$ to
$-21.4$~mag. The majority of the upper limits on non-detected hosts are as
deep, and deeper, than any of the host detections, particularly at high
redshift ($z> 3$). All upper limits, except for one host at $z=6.295$, are well
below the median magnitude of the detected sample.

It is interesting to note that the brightest hosts, i.e., those above the median
absolute magnitude, span a quite limited redshift range between $z\sim1$ and
$z\lesssim3$. Conversely the dimmest 50\% of hosts (detections and upper limits)
span the entire redshift range of the sample. This evolution is similar to that
of the UV-inferred global star formation-rate density \citep[e.g.,][]{Daddi2007a,
Noeske2007a, Rodighiero2010a, Elbaz2011a, Bouwens2014b}, which also peaks at
$z=2$--3.

\begin{figure}[t!]
\begin{center}
\includegraphics[width=1\columnwidth]{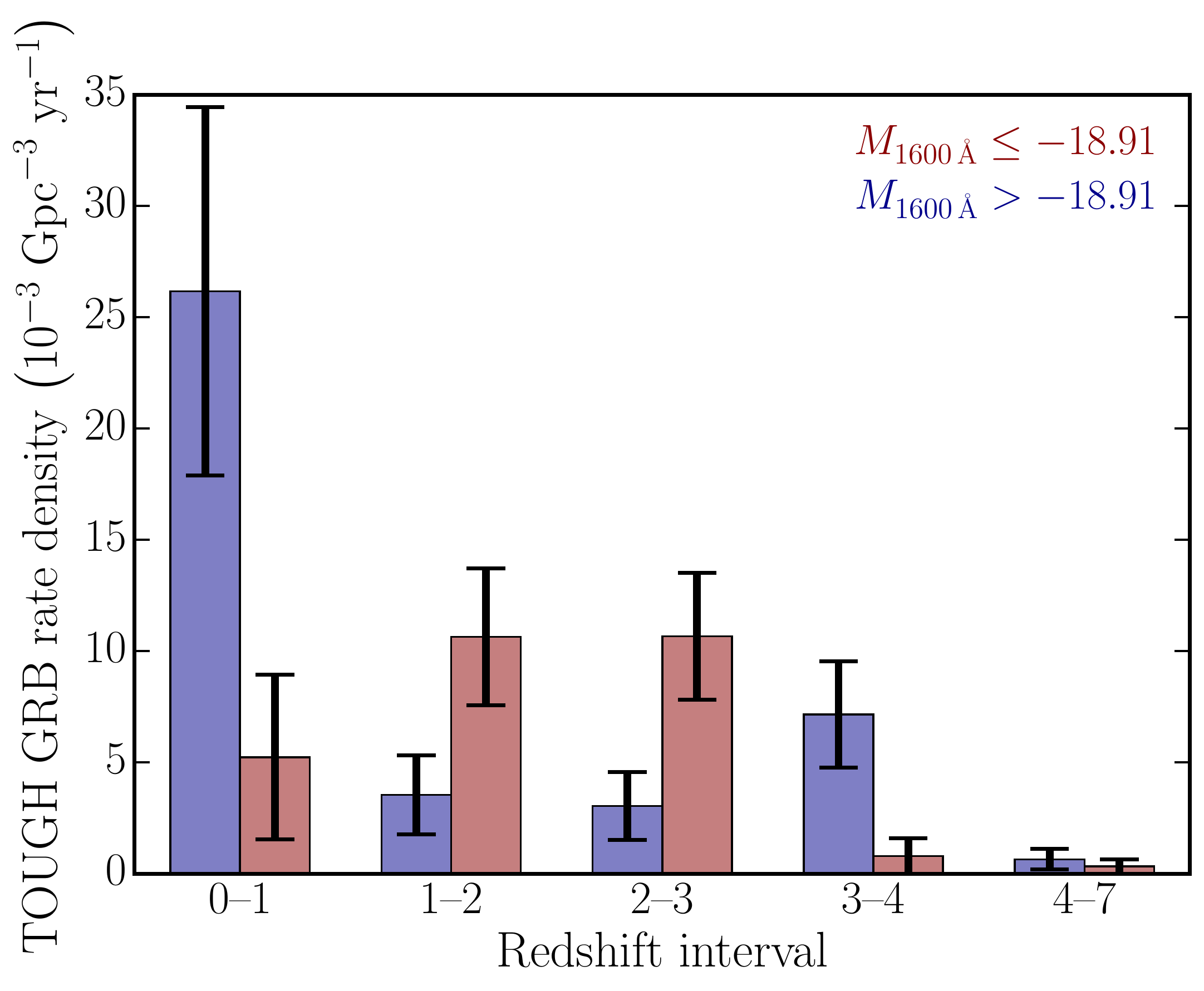}
\caption{
	GRB rate density (normalized to the TOUGH period of 2.5 years; \citealt{Hjorth2012a})
	in hosts below and above the median UV luminosity
	of the TOUGH sample. For simplicity only hosts with known redshifts are shown.
	GRBs occur in faint hosts at $z<1$ about a factor of 5 times more frequently,
	while at $z=1$--3, bright hosts are factor of 2.5--3 times more common more
	common than faint hosts. Error bars represent Poisson errors on each bin. For
	details see Section \ref{sec:results_lum}.
}
\label{fig:GRBD}
\end{center}
\end{figure}

Above $z>3$, there is a conspicuous absence of any host detection above the
median magnitude, except for the single bright $M_{1600\,\textrm{\AA}}=-20.8$~mag
host (GRB\,060707) at $z=3.424$, though the host of GRB050922B could have similar
luminosity (For details see Section \ref{sec:photoz}). At the other end of the
redshift scale, below $z\sim1$, there are once again no hosts above the median
magnitude. Our unbiased host sample seems to suggest that GRBs favor lower
luminosity hosts throughout the entire redshift range, in which they are detected,
and are only found in UV brighter hosts in the range between $1<z<3$, though
we note (as discussed below) that some of the hosts with unknown redshifts may
in reality exist in the higher redshift range. 

Figure \ref{fig:GRBD}
plots the rate density (number per unit
comoving volume per year) of TOUGH GRBs occurring in hosts above and below the
overall survey median luminosity (dotted line left panel Figure \ref{fig:scatter_1})
versus redshift. Although the overall numbers are small, the volume density of the
brighter host fraction is lower than that of the fainter fraction at all redshifts
except $1<z<3$ where bright hosts reverse this trend to become 2.5--3 times more
common than those below the global median. We caution that the relatively low numbers
in each large and somewhat arbitrary redshift bin prevent any firmer statistical
conclusion being drawn from the binned data. Further analysis of the cumulative
distribution function (CDF) of the TOUGH hosts compared to model LFs
derived from LBG populations in Sect \ref{ref:cdf} investigates these preliminary
observations in greater detail.

In the following sections, we investigate how GRB hosts trace the SFR history,
in particular whether they follow the model proposed by \citet[][see also
\citealt{Jakobsson2005a}]{Fynbo2002a} of GRBs selecting galaxies from a general
population according to their SFR, or whether an additional dependency must be
invoked, such as metallicity \citep{Stanek2006a,Modjaz2008a,Levesque2010a}.

We note that the redshifts of nine hosts are uncertain (Table\ \ref{tab:host_mag})
within defined limits \citep[for details see][]{Jakobsson2012a}, one host
identification (GRB 060805A) is non-unique and the IGM correction for one host
(GRB 050502B) is uncertain. In the right panel of Figure\,\ref{fig:scatter_1} we
plot (gray lines for hosts of uncertain $z$, green data points for the host with
uncertain IGM correction, and blue data points for the non-unique host
identification) their possible positions in the $M_{1600\,\textrm{\AA}}$-$z$
plane, where it can be seen that the plausible distribution of these uncertain
hosts shows no discrepancy from the firm detections/limits. We discuss further
the effects of these unknown redshift hosts on our results in Section
\ref{sec:unknown_z}.

\begin{figure}[t!]
\begin{center}
\includegraphics[width=1.0\columnwidth]{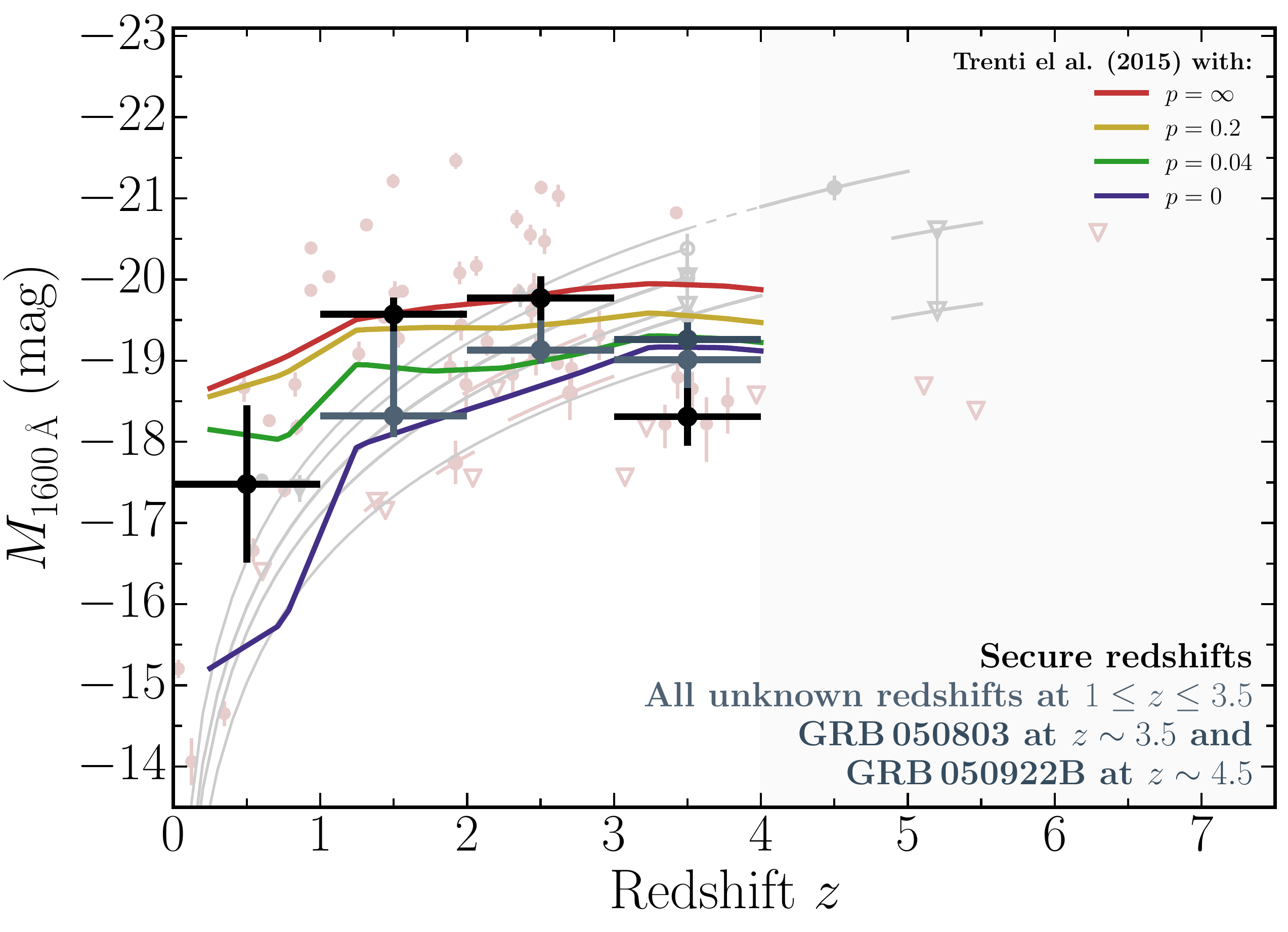}
\caption{
	Evolution of the median obscured UV luminosity (in bins of unit redshift)
	with known redshift (black markers). Their errors were assessed through
	bootstrapping. At $z=3$--4 the sample is characterized by a significant
	number of non-detections. Overlaid are model tracks for different
	strengths of a possible metallicity bias \citep{Trenti2014a}. The hosts
	with unknown redshift were also put at $1<z<2$ and $2<z<3$ to assess
	their impact on the median value (displayed by lighter blue bars). The
	blue bar shows the median UV luminosity if GRBs\, 050803 and 050922B
	are put at $z=3.5$ and $z=4.5$, respectively (Section \ref{sec:photoz}).
	See Section \ref{sec:results_Zbias} for details.
	}
\label{fig:Trenti}
\end{center}
\end{figure}

\subsection{The Evolution of the Median UV luminosity}\label{sec:results_Zbias}\label{sec:results_sfrd}

As a first diagnostic to investigate whether GRB are biased or unbiased tracers
of star formation, we investigate the evolution of the median obscured UV
luminosity of the hosts. Recently \citet{Trenti2014a} presented tracks for
the evolution of the median observed UV luminosity of GRB host galaxies for
various levels of GRB production bias with respect to host metallicity,
characterized by a parameter $p$ which represents a minimum, metal-independent
plateau for the efficiency of forming a GRB. Thus a low value of $p$ characterizes
a high level of bias toward low metallicity hosts ($p=0$ representing a stringent
metallicity cutoff), whereas $p\rightarrow\infty$ characterizes no metallicity
bias.

Figure \ref{fig:Trenti} shows the evolution of the median UV luminosity in unit
redshift bins of our sample in comparison to these models. The luminosity depth
of our observations is a function of redshift (Figure \ref{fig:scatter_1}), and we
hence recalculated the median values given in \citet{Trenti2014a} for the
observed luminosity limits in each redshift bin. The faint limit was set to
$M-3\sigma(M)$ in each magnitude bin; specifically we compute the median UV
luminosity between $-22.5\leq M\leq-13.2$ at $z<1$, $-22.5\leq M\leq-16.7$ at
$1\leq z<3$, $-22.5\leq M\leq-17.2$ at $3\leq z < 4.5$. Observed medians and
their errors were estimated via a bootstrap method (30,000 samples), where each
detected host was represented as a Gaussian centered on the measured luminosity
with a width given by the measurement uncertainty. Non-detected hosts were drawn
from a uniform distribution between a fiducial magnitude cut and the $3\sigma$
limiting detection.

Though we have only four data points with significant uncertainty, the data
appear to favor models of GRB production incorporating significant levels of
bias toward low metallicity hosts \citep[as suggested by][]{Vergani2014a, Cucchiara2015a}.
In particular, the inclusion of any of the hosts of unknown redshift in
individual redshift bins always lowers the median magnitude in that bin. Note
that these models of metallicity bias for GRB production, which include a metallicity
dependent (single star collapsars) and independent channel (binary progenitor)\footnote{\citet{Cantiello2007a}
argue that the binary channel may prefer low-metallicity environments as well, i.e.,
being metal dependent.} do not imply a fixed fraction of GRB production
via each channel, but rather an evolving fraction with redshift, with the
metallicity-dependent (collapsar type) channel always being dominant at high
redshift (see \citealt{Trenti2014a} for full details).

\begin{figure}
\begin{center}
\includegraphics[width=1.0\columnwidth]{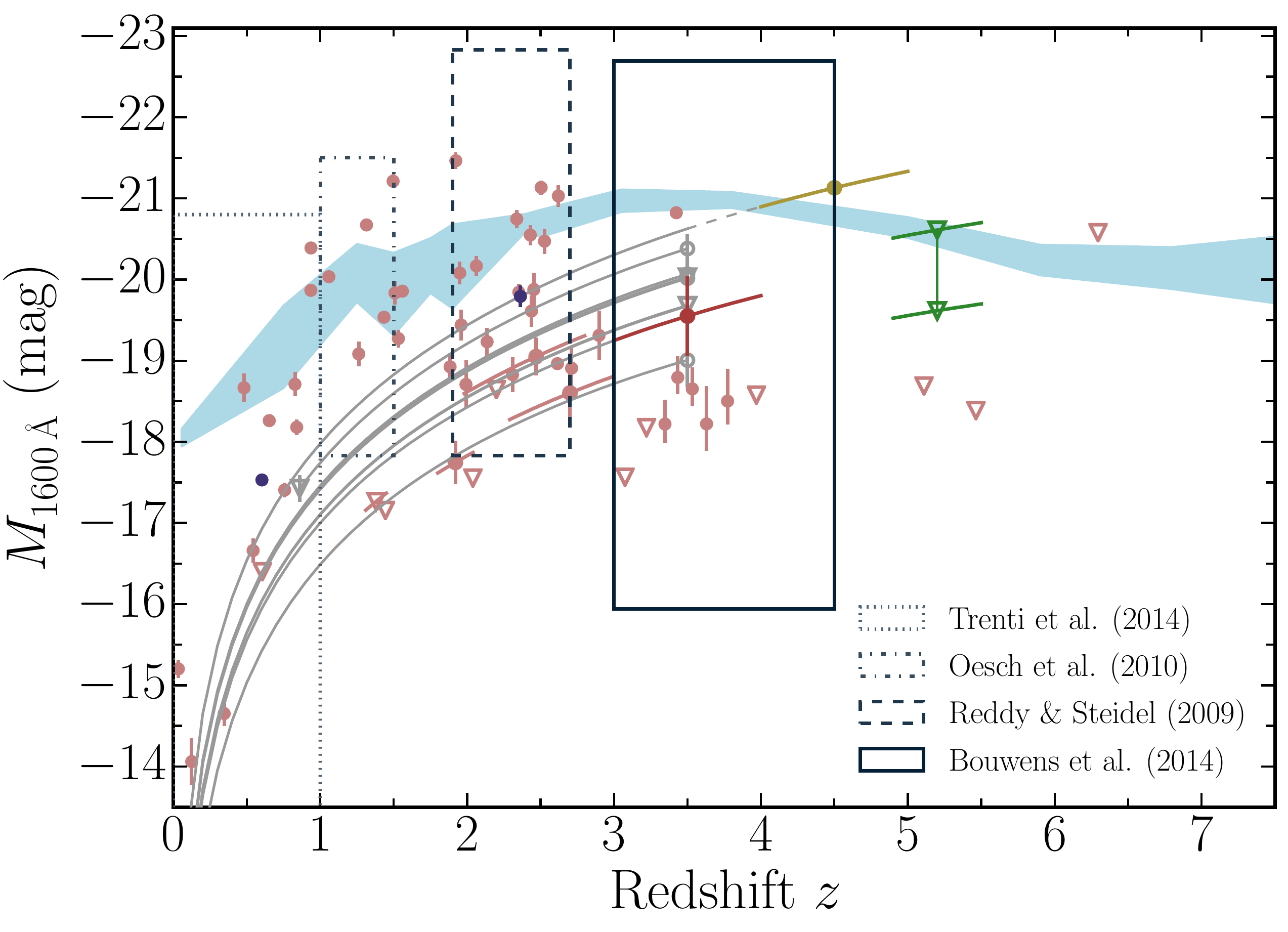}
\caption{
	Observed luminosity evolution of GRB host galaxies and the evolution of
	$M_{1600\,{\textrm \AA},\star}$ of LBGs (light-blue shaded area; \citealt{Arnouts2005a};
	\citealt{Reddy2009a}; \citealt{Oesch2010a}; \citealt{Bouwens2014b}).
	Overlaid are the redshift and luminosity ranges probed by the
	LBG surveys in Table\ \ref{tab:lf_lit} at $z>1$ and the predicted LBG LF
	for $z<1$ by \citet{Trenti2014a}, which are used to construct the UV
	GRB-host luminosity function. The color coding
	is identical to Figure \ref{fig:scatter_1}.
	}
\label{fig:scatter_3}
\end{center}
\end{figure}

\subsection{The UV Luminosity Function}\label{ref:cdf}

In order to go further than just analyzing the median properties of the TOUGH
hosts, we construct luminosity functions for GRB hosts from $z=0$ to $z=4.5$ in
appropriate redshift and luminosity bins to compare both with those from LBG surveys
and those predicted by the models of \citet{Trenti2014a}. 
We select those hosts from our sample that fall in the redshift ranges of the
three LBGs surveys in Table\ \ref{tab:lf_lit}, and in the range $0<z<1$.

Figure\,\ref{fig:scatter_3} shows which part of the $M_{1600\,\textrm{\AA}}$-$z$
parameter space is probed by LBG surveys. The evolution of $M_\star$ implies
that GRB hosts probe the full luminosity range of LBGs between $z=1$ and 3,
whereas at lower and higher redshifts, GRBs rather probe the faint-end of the
observed LBG luminosity function. We note that the luminosity range of GRB host
galaxies extends to much fainter galaxies between $z=1$ and $z=3$ than probed by
the LBG surveys in Table \ref{tab:lf_lit}. 
However, recent observations by \citet{Alavi2014a} found no evidence for a change
in the LBG luminosity function parameters at $z\sim2$ down to $M_{1500}=-15$~mag, 
which reassures us in extrapolating the luminosity functions in Table \ref{tab:lf_lit}
to lower luminosities.

\begin{figure}
\begin{center}
\includegraphics[width=1.0\columnwidth]{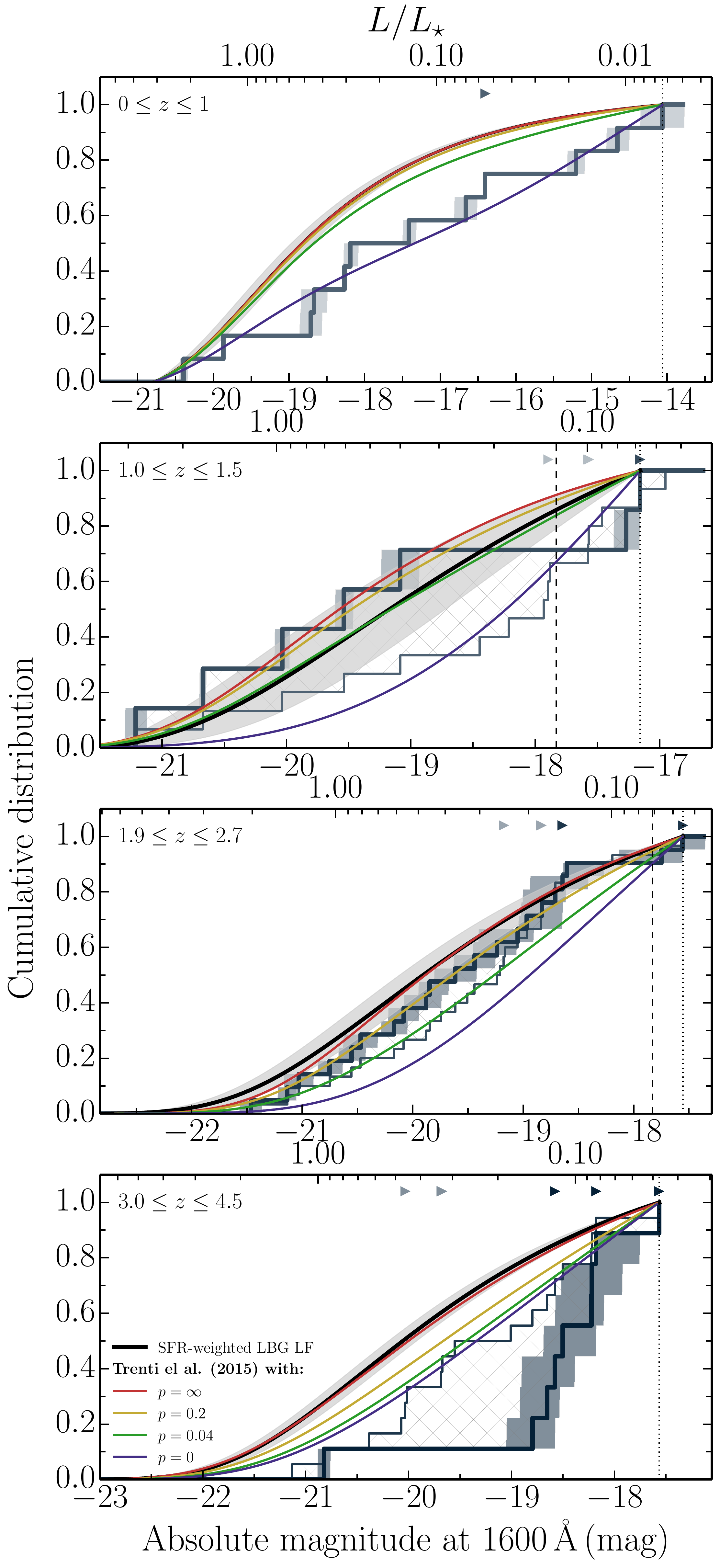}
\caption{The observed cumulative distribution (CDF) of the TOUGH sample at different
	redshifts. The shaded areas display the $1\sigma$ uncertainties
	in the UV luminosity of the GRB host galaxies. The hatched regions
	shows the parameter space between the two extreme cases of
	including/not including all hosts of unknown redshift in the
	respective redshift interval. The black curves display the CDFs
	of a SFR-weighted LBG luminosity function for the given redshift
	interval and their $1\sigma$ envelopes in gray.
	In color are shown the luminosity functions for various levels of
	metallicity bias predicted by \citet{Trenti2014a}. The limiting
	magnitudes of the LBG surveys are displayed by the dashed vertical
	lines and the dotted vertical lines indicate the magnitude of the faintest
	observed host in each sample, where the CDFs are formally
	normalized in each respective panel. The parameters of the LBG luminosity
	functions are listed in Table\ \ref{tab:lf_lit}.
}
\label{fig:cdf}
\end{center}
\end{figure}

\begin{table}
\caption{Parameters of LBG Luminosity Functions}
\centering
\scriptsize
\begin{tabular}{c@{\hspace{1mm}}c@{\hspace{1mm}}c@{\hspace{1mm}}c@{\hspace{1mm}}c}
\toprule
Redshift		& Luminosity 		& $ M_{1600\,\textrm{\AA},\star}$	& Faint-end 	& \multicolumn{1}{c}{\multirow{2}*{Ref.}}	\\
Interval		& Interval (mag)	& (mag)					& Slope $\alpha$&						\\
\midrule
$1.0\leq z\leq 1.5$	& $-21.50$ to $-17.83$	& $-20.08\pm0.36$			& $-1.84\pm0.15$			& (1)\\[0.5mm]
$1.9\leq z\leq 2.7$	& $-22.83$ to $-17.83$	& $-20.70\pm0.11$			& $-1.73\pm0.07$			& (2)\\[0.5mm]
$3.0\leq z\leq 4.5$	& $-22.69$ to $-15.94$	& $-21.07\pm0.08$			& $-1.64\pm0.04$			& (3)\\[0.5mm]
\bottomrule
\end{tabular}
\tablecomments{Values of LBG LFs in the Schechter parametrization for
	different redshift intervals. The redshift column shows the interval for which the
	luminosity functions were applied to.
}
\tablerefs{(1) \citet{Oesch2010a}, (2) \citet{Reddy2009a}, (3) \citet{Bouwens2014b}}
\label{tab:lf_lit}
\end{table}

Figure\,\ref{fig:cdf} displays the GRB host galaxy cumulative distributions for
the four redshift intervals. Since GRBs are produced by the collapse of very
massive stars, it has been suggested that GRBs should select galaxies according
to their SFR. In the simplest model, \citet{Fynbo2002a} proposed that the UV
GRB-host luminosity function should be similar to that LBG samples weighted by
their SFR, where the SFR is proportional to the \textit{unobscured}
UV luminosity, which can be expressed
as ${\rm SFR}\propto 10^{0.4\times A_V(M_{\rm obs})}\,L_{\rm obs}$. Following
\citet{Trenti2014a}, we use
$A_V=4.43 + 0.79\,\log(10)\,\sigma^2 _{\beta_{\rm UV}}+1.99\,\beta_{\rm UV}$
where $\sigma_{\beta_{\rm UV}}=0.34$.
The cumulative distributions of the SFR-weighted LBG luminosity functions from
Table \ref{tab:lf_lit} are overlaid in the same plot.

At low and high redshift, the observed distribution differs significantly from
the no metallicity bias (or equivalently, LBG derived) model LFs. In the 
$3<z<4.5$ region for example, we can reject the null hypothesis
of the data being drawn from the SFR weighted LBG LF at a level of $0.01\%$ via a KS test.
In the medium redshift ranges ($1<z<1.5$ and $1.9<z<2.7$) the data are
insufficient to distinguish reliably between any of the models, particularly due 
to some non-detections and the possibility of some of the hosts of unknown
redshift being within these bins as discussed in the next section. 

\begin{figure}
\begin{center}
\includegraphics[width=1.0\columnwidth]{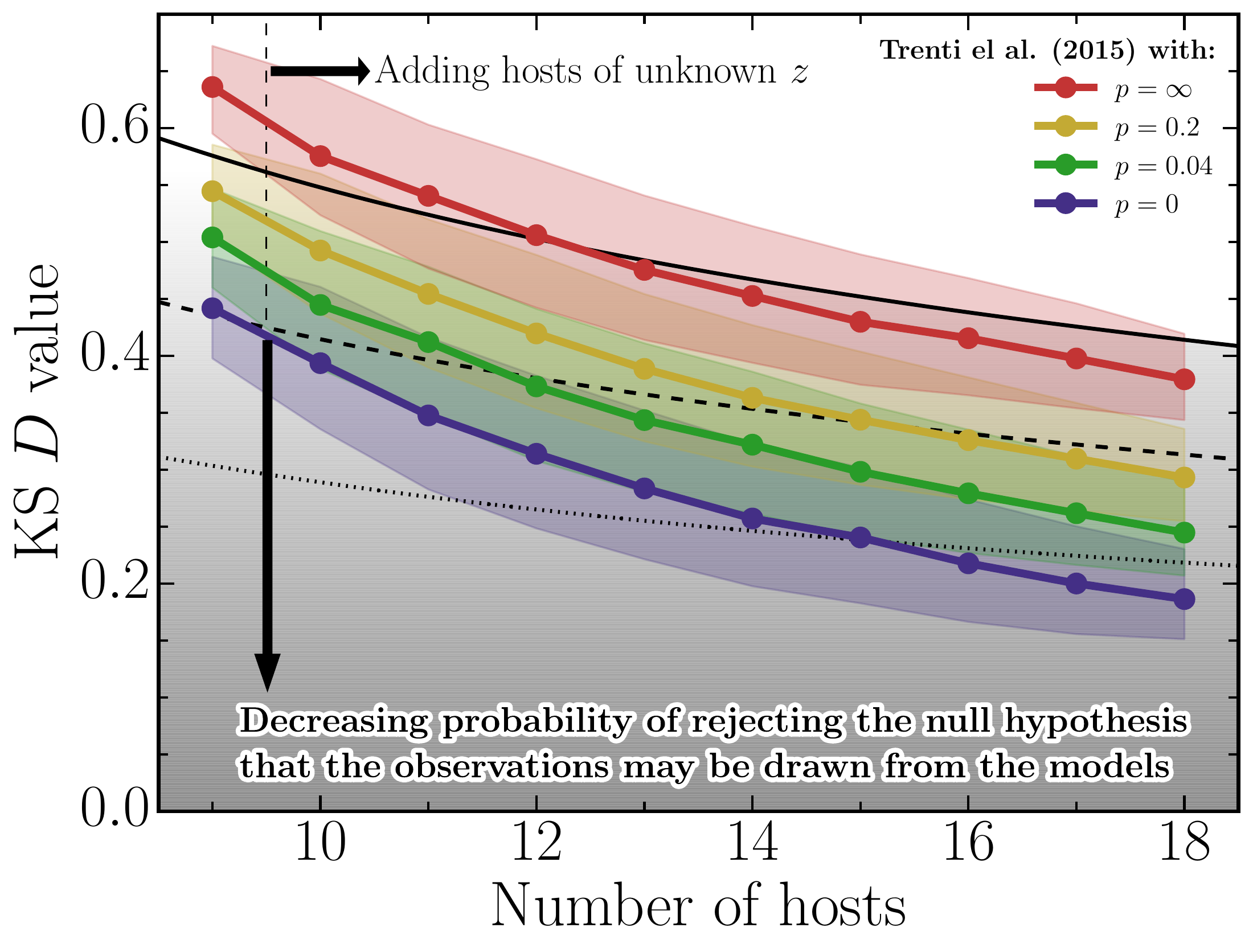}
\caption{Kolmogorov-Smirnov test between the observed TOUGH hosts and the
	 various luminosity functions between $z=3.0$ and $z=4.5$ as the
	 number of hosts with unknown redshift is increased. The color-shaded
	 regions represent the 16 and 84 percentiles of the MC simulations.
	 With the null hypothesis that the data may be drawn from the
	 respective model LF, the white area of the plot represents rejection
	 of this null hypothesis with a probability level of it being correct of
	 $<0.3\%$ ($3\sigma$ equivalent). Within the gray shaded region,
	 the dashed and dotted lines display the $5\%$ ($2\sigma$ equivalent)
	 and $32\%$ ($1\sigma$ equivalent) rejection levels respectively.
	 See Section \ref{sec:unknown_z} for details.
}
\label{fig:ks_2}
\end{center}
\end{figure}

\section{Discussion}\label{sec:discussion}
\subsection{The Impact of Hosts with Unknown Redshifts}\label{sec:unknown_z}

There are nine hosts in our sample with redshifts unknown within certain limits.
As discussed in \cite{Jakobsson2012a}, a conservative upper limit of $z<3.5$ can
be placed on four of the bursts associated with these hosts via their measured
excess (above Galactic) X-ray absorption following the method of \citet{Grupe2007a}.
The remaining five bursts can be at higher redshifts (see Table \ref{tab:host_mag}
for details).

In this paper we set a formal redshift limit of $z<3.5$ to avoid
corrections for IGM and Ly$\alpha$ absorption. These unknown redshift hosts can
therefore exist anywhere along the gray tracks plotted in the right hand panel
of Figure\,\ref{fig:scatter_1}. The maximum plausible redshift of $z=3.5$ is
approximately coincident with the midpoint of our highest redshift LBG comparison
sample, and in Figure\ \ref{fig:cdf} we also show the effect of placing all the
unknown hosts at the midpoint of each LBG comparison range in turn (hatched
regions). As can be seen from the bottom panel in this plot, if all unknown
redshift hosts are actually at their highest plausible redshift, which
may be a likely scenario (see Figure\,11 in \citealt{Hjorth2012a}), and all upper
limits are treated as detections, then the likelihood (KS probability) of the
TOUGH cumulative distribution being consistent with the SFR weighted LBG model
rises to 1.0\%. Note that this value represents our most conservative
limit compared to the more flexible simulation discussed below.

To quantify how many of the TOUGH hosts with unknown redshift would be required
to be within the range $3\leq z\leq 3.5$ in order to make the TOUGH host
luminosity function consistent with the SFR-weighted LBG LF, we performed a
Monte-Carlo simulation as follows. One of the nine hosts with unknown redshift
is chosen at random and assigned a random redshift between $3\leq z\leq 3.5$.
The appropriate host luminosity is then drawn from a normal distribution centered
around the observation value with a width ($1\sigma$) of the detection error for
detected hosts, and a uniform distribution between the upper limit and the
luminosity of the faintest host in the TOUGH $3\leq z\leq 4.5$ sample for those
hosts with an upper limit only. This host is added to the TOUGH CDF, and a KS value
computed between the new CDF and the SFR weighted LBG-LF. The process is repeated
30,000 times and a mean and median KS value obtained. We then successively add
hosts of unknown redshift and repeat the procedure until all nine hosts are
included in the CDF. Figure \ref{fig:ks_2} shows how the measured KS value varies
with the number of added hosts of unknown redshift.

As can be seen from Figure \ref{fig:ks_2}, at least 3 unknown hosts are
required for the TOUGH distribution to fall below the $3\sigma$
equivalent rejection probability level of $<0.3\%$
chance of consistency with being drawn from the SFR-weighted LBG
(no metallicity bias) LF model. Even when all unknown redshift
hosts are included in the simulation, the median KS probability of the
distributions being consistent is only 0.8\%. The gap may be closed further if
some of the unknown hosts are at higher redshifts still ($3.5<z<4.5$) since
their $R$-band observations then would have been significantly affected by IGM and
Ly$\alpha$ absorption.  Nevertheless, the metallicity dependent models
of \citet{Trenti2014a} appear a better fit to the
data, with critical KS rejection levels much lower.

\begin{figure*}
\begin{center}
\includegraphics[width=1.0\textwidth]{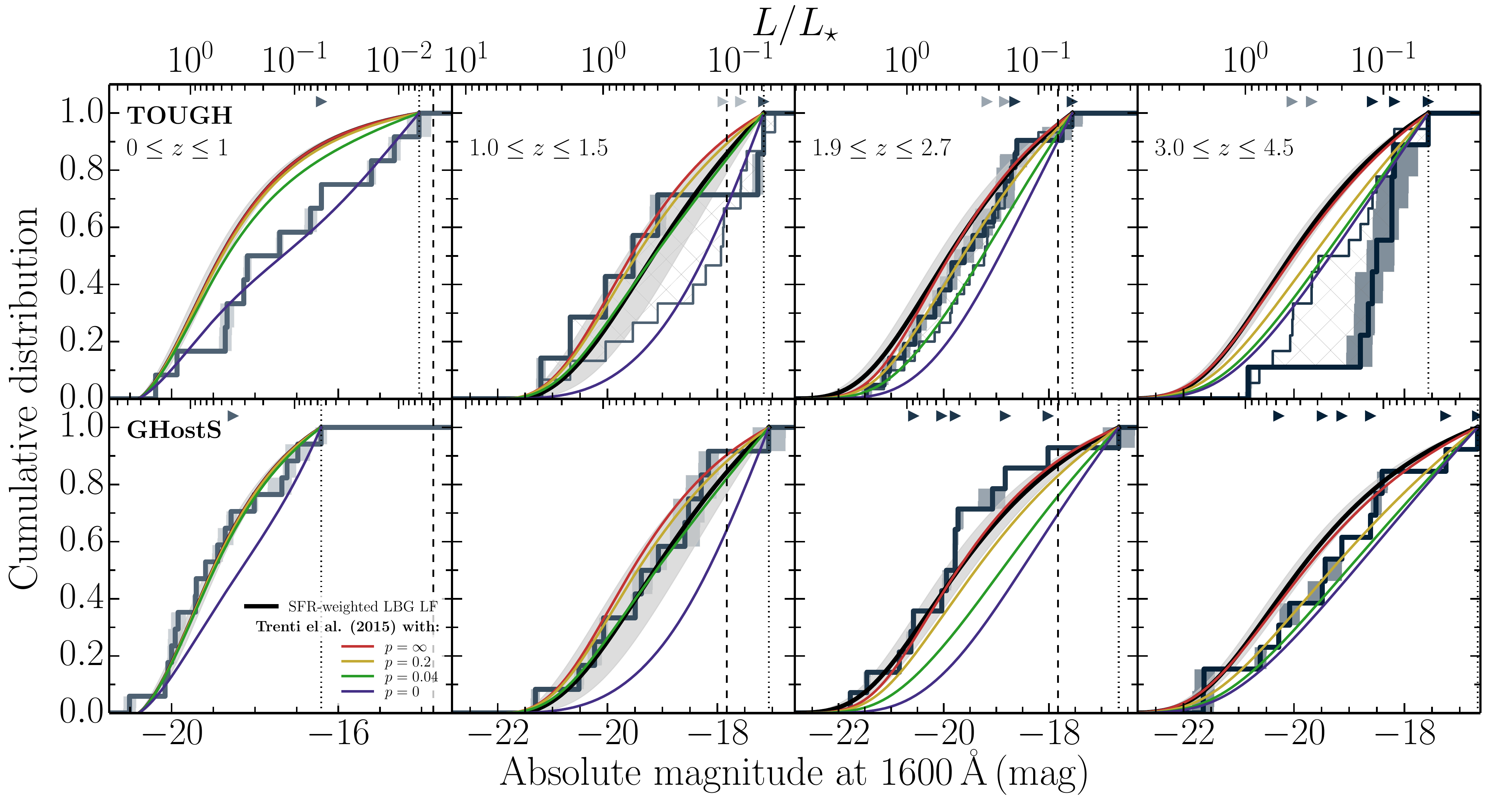}
\caption{The observed CDFs
	of the heterogeneous GHostS (bottom) sample in comparison to the
	optically unbiased TOUGH (top) sample at different redshifts. There are
	clear differences between the two host samples, particularly at
	high-$z$ (see text for discussion). The hatched regions shows the
	parameter space between the two extreme cases of including/not including
 	all hosts of unknown redshift in the respective redshift interval. The 
	black curves display the CDFs of a SFR-weighted LBG luminosity function
	for the given redshift interval and their $1\sigma$ envelopes in gray.
	In color are shown the luminosity functions for various levels of
	metallicity bias predicted by \citet{Trenti2014a}. The plot boundaries
	were adjusted to the luminosity range of the respective LBG survey and
	\citet{Trenti2014a}. The thick line shown in light blue in the
	$3\leq z \leq 4.5$ panel indicates the luminosity function, if all hosts
	with unknown redshift would be at $z=3.5$, and GRBs\, 050805 and 050922B
	at $z=3.5$ and $z=4.5$, respectively. The limiting
	magnitudes of the LBG surveys are displayed by the dashed vertical
	lines and the dotted vertical lines indicate the magnitude of the faintest
	observed host in each sample, where the CDFs are formally
	normalized in each respective panel. The parameters of the LBG luminosity
	functions are listed in Table\ \ref{tab:lf_lit}.
	}
\label{fig:cdf2}
\end{center}
\end{figure*}

We emphasize again that these models do not represent a fixed fraction of GRB
production via metal dependent and independent channels, but are derived from
stellar population syntheses combined with a GRB production efficiency model
with respect to metallicity controlled by the value of $p$. Lower (but non-zero) $p$ values
represent a stronger level of bias toward low metallicity hosts, as expected
from the collapsar model. We note that the
$p=0$ model (blue lines) which represents a stringent metallicity plateau for
GRB production is effectively ruled out by observations of some high metallicity GRB
hosts \citep[e.g.,][]{Levesque2010a, Elliott2013a, Schulze2014a}, and are plotted
as convenient limits for comparison only.

\subsection{Influence of Selection Effects on Previous GRB Host Samples}\label{sec:selection_effects}

The TOUGH sample is independent of the brightness of the optical afterglow, but
the properties of the prompt emission could still bias the sample toward a
certain GRB population and hence a certain host galaxy population. We find no
correlation between the host properties and energy released at $\gamma$-ray
energies, $E_{\rm iso}$ (values taken from \citealt{Butler2007a}), or the peak
flux photon flux (taken from \citealt{Sakamoto2008a}) at any redshift. At low
redshift (where we are sampling the greatest proportion of the GRB $\gamma$-ray
luminosity function itself) there is also no tendency for brighter bursts to favor
dimmer hosts, and we therefore do not believe any flux limited observational
bias of the bursts themselves significantly affects any of our conclusions
regarding the hosts.

Previous GRB host surveys have tended to be biased toward bursts with optically
brighter afterglows and/or hosts, whereas the TOUGH sample (as stressed throughout
this work) is independent of the optical properties of bursts or hosts. To
examine whether these previous observational biases are significant or not, we
investigate whether the same conclusions could be drawn from a previous
heterogeneous sample. To address this question we retrieved all host measurements
that probe the rest-frame UV continuum from the GHostS database
\citep{Savaglio2006a}. Until 2012 January (before the first results from the
TOUGH survey were published), the database comprised observations for over 120
hosts. As an example, we build the UV luminosity distributions for this sample
and divide them into the same redshift and luminosity bins as the TOUGH sample.

The observed cumulative distributions are shown in the lower panels of
Figure\ \ref{fig:cdf2}, along with those of the TOUGH sample in the upper panels.
At all redshifts, except perhaps for $1<z<1.5$, we see an overabundance
of luminous galaxies in the GHostS sample compared to TOUGH. This clearly shows
that selection effects from GRB afterglow observations must be carefully taken
into account before attempting to extract any ensemble properties. The limiting
magnitudes in the GHostS sample are in general also significantly shallower
than ours. To be able to extract any meaningful results, GRB-selected galaxy
samples need to be carried out to the magnitude limits of LBG surveys, i.e.,
reaching at least $R\sim28$~mag.

\subsection{Origin of the Redshift Evolution of the Luminosity Function}

Our findings let us draw the following picture. GRBs probe the full luminosity
range (and below) of LBG surveys at all comparable redshifts. Furthermore,
throughout the entire redshift range of our sample, GRB hosts appear consistent
with models of GRBs being produced by a combination of a metal-dependent (single
star collapsar) and independent (binary progenitor) channel with a relatively
high level of bias (low, but non-zero value of $p$) toward low metallicity
hosts \citep{Trenti2014a}, though we cannot completely rule out the unbiased
models in the range $1<z<3$. This behavior may also lend support to the model
of the starvation of infalling pristine gas at $z\sim1$–-1.5 proposed by
\citet{Perley2013a}, moving the GRB host population away from LBG galaxies as
a whole to low-mass UV faint galaxies as redshift decreases.

At $z>3$, we observe a lack of UV luminous GRB hosts. All detected hosts (except
one) are at least 3 mag fainter than $L_\star$ galaxies. If GRB hosts
were truly similar to LBGs, we may expect this connection to improve at higher
redshifts, where any potential metallicity bias becomes less important in
comparison to the overall galaxy population. Indeed the presence of a reasonable
fraction of the nine hosts of unknown redshift in this range would reduce this
lack of bright hosts, though not entirely to the level of consistency with an
SFR weighted LBG LF.

Thus it would seem that at both high and low redshift, TOUGH supports the idea
that GRBs are preferentially found in low-metallicity, relatively UV faint hosts
compared to the overall galaxy population, and therefore their relationship to
the cosmic SFR density history as measured by UV flux limited surveys is not
simple. Several previous studies have pointed to the numbers of GRBs observed at
high redshift indicating a large amount of hidden (massive) star formation in
galaxies that are too faint to be detected in current LBG surveys \citep{Fynbo1999a,
Haehnelt2000a, Fynbo2002a, LeFloch2003a, Jakobsson2005a, Kistler2008a,
Schmidt2009a, Jakobsson2012a}.

Alternatively others have suggested that GRB production efficiency with respect
to star formation may increase with redshift \citep{Kistler2009a, Butler2010a,
Robertson2012a, Salvaterra2012a}. If this latter alternative of increased GRB
efficiency were the sole reason for the change however, one might expect to
still see some bright hosts at high redshift. However, highly star-forming
galaxies may well also contain a large amount of dust, both obscuring their
intrinsic SFR and making it more difficult to obtain spectroscopic
redshifts. Despite targeted and deep searches, no GRB host has
yet been detected beyond $z\sim5$, which lends support to the lack of bright
hosts at high redshift not being an artifact of our survey. Obviously since high-$z$ GRBs
are rare, the influence of small number statistics cannot be avoided. We suggest
however, that though an increase in GRB production efficiency with respect to the UV
measured cosmic SFR density at high redshift may be a useful calculational tool
in simulation studies, it does not represent the detailed picture regarding
individual GRB host galaxies, and more unbiased GRB host data are required.

During the review process of this paper, other workers have submitted investigations of
the properties of GRB host galaxies. \citet{Greiner2015a} presented discussions of the
properties of a \textit{heterogeneous} sample of GRB hosts at $z\sim3$, and \citet{Perley2015a}
presented findings on the NIR luminosity distribution $z = 0$ to $z = 5$ from the optically
unbiased \swift\ Gamma-Ray Burst Host Galaxy Legacy Survey (SHOALS; \citealt{Perley2015b}).
While the former study concludes that GRBs are unbiased tracers of star formation (compare
our Section \ref{sec:selection_effects} and Figure \ref{fig:cdf2}), the latter study comes to a similar
conclusion to that presented in this paper: GRB hosts in general appear fainter than a model
that solely depends on SFR would predict, with a particularly pronounced lack of NIR luminous
galaxies at $z < 1$.
In all three samples, the
number of events at $z>4$ is relatively small, illustrating the need for a more concerted
observational effort targeting this regime.

\section{Conclusion}\label{sec:conclusion}

We have used the optically-unbiased TOUGH sample to examine the UV GRB
host galaxy LF from $z=0$ to $z=6.3$. We find the TOUGH host
LF to be most compatible at all redshifts with an SFR weighted
LF derived from a model containing both a metal-independent
(binary progenitor) and metal-dependent (single star collapsar) channel with a 
relatively high level of bias toward low-metallicity hosts as suggested in 
\citet{Kocevski2011a}, \citet{Graham2013a}, \citet{Perley2013a}, \citet{Vergani2014a}, \citet{Cucchiara2015a}, and \citet{Trenti2014a}. This is particularly
the case at low ($0<z<1$) and high ($3<z<4.5$) redshifts, though we cannot rule out an
unbiased LF at medium redshifts ($1<z<3$) as observed in more heterogeneous
samples with unknown observational biases.

At high redshifts ($3<z<4$) in particular, the lack of detected UV luminous host
galaxies combined with the likelihood that several (if not all) of the nine hosts
of unknown redshifts may be in this range show that a deep spectroscopic survey
with an optical+NIR spectrograph such as the ESO X-shooter UV--NIR echellete
spectrograph is needed to fully constrain the GRB host LF and elucidate more
fully the relationship between GRBs and the cosmic SFR density history.

---------------------------------------------------------------------
\section*{Acknowledgments}

\textit{We acknowledge with sadness the recent unexpected passing of our colleague
and co-author Javier Gorosabel. His support of and contributions to this work and
astronomy in general are greatly appreciated.}
We thank the referee for a careful reading of the manuscript and for helpful comments
that improved this paper. We thank Stephanie Courty and Daniele Malesani for fruitful
discussions. We also thank Daniele Malesani for kindly providing the TNG data, and Michele
Trenti for providing the parametrization of the spectral slope.
S. Schulze acknowledges support from CONICYT-Chile FONDECYT 3140534, Basal-CATA
PFB-06/2007, and Project IC120009 ``Millennium Institute of Astrophysics (MAS)´´
of Iniciativa Cient\'{\i}fica Milenio del Ministerio de Econom\'{\i}a, Fomento
y Turismo, by a Grant of Excellence from the Icelandic Research Fund, and from
the University of Iceland Research Fund.
RC is grateful to the University of Hertfordshire for travel support.
The research activity of AdUP and JG is supported by Spanish research project
AYA2012-39362-C02-02. AdUP acknowledges support by the European Commission under
the Marie Curie Career Integration Grant program (FP7-PEOPLE-2012-CIG 322307).
The Dark Cosmology Centre is funded by the Danish National Research Foundation.
The research leading to these results has received funding from the European
Research Council under the European Union's Seventh Framework Program
(FP7/2007-2013)/ERC Grant agreement no. EGGS-278202.
This research has made use of the GHostS database
(\href{www.grbhosts.org}{www.grbhosts.org}), which is partly funded by
\textit{Spitzer}/NASA grant RSA Agreement No. 1287913. Based in part on
observations collected
with the NASA/ESA \textit{Hubble Space Telescope}, obtained at the Space Telescope
Science Institute, which is operated by the Association of Universities for Research
in Astronomy, Inc., under NASA contract NAS 5-26555, as part of the program 11734,
12307, 11734,
with the Gemini Observatory, which is operated by the Association of Universities for
Research in Astronomy, Inc., under a co-operative agreement with the NSF on behalf of
the Gemini partnership, as part of the programs GS-2007B-Q-1, GS-2013B-Q-69, and
GS-2014A-Q-6,
the Gran Telescopio Canarias (GTC), installed in the Spanish Observatorio del Roque
de los Muchachos of the Instituto de Astrof\'{i}sica de Canarias, in the island of La
Palma,
with the Italian Telescopio Nazionale Galileo (TNG) operated on the island of La Palma
by the Fundaci\'on Galileo Galilei of the INAF (Istituto Nazionale di Astrofisica) at
the Spanish Observatorio del Roque de los Muchachos of the Instituto de Astrofisica de
Canarias as part of the programs A22TAC-107, with the 2.2-m MPG telescope on La Silla 
as part of the program CN2014B-102, and
with the Spitzer Space Telescope, which is operated by the Jet Propulsion Laboratory,
California Institute of Technology under a contract with NASA, as part of the program
57753,
Some of the data presented herein were obtained at the W.M. Keck Observatory, which is
operated as a scientific partnership among the California Institute of
Technology, the University of California and the National Aeronautics and
Space Administration. The observatory was made possible by the generous
financial support of the W.M. Keck Foundation.
Part of the funding for GROND was generously granted from the Leibniz-Prize to Prof.
G. Hasinger (DFG grant HA 1850/28-1).
Funding for SDSS-III has been provided by the Alfred P. Sloan Foundation, the
Participating Institutions, the National Science Foundation, and the U.S. Department
of Energy Office of Science. The SDSS-III web site is \href{http://www.sdss3.org/}{http://www.sdss3.org/}.
SDSS-III is managed by the Astrophysical Research Consortium for the Participating
Institutions of the SDSS-III Collaboration including the University of Arizona,
the Brazilian Participation Group, Brookhaven National Laboratory, Carnegie Mellon
University, University of Florida, the French Participation Group, the German
Participation Group, Harvard University, the Instituto de Astrofisica de Canarias,
the Michigan State/Notre Dame/JINA Participation Group, Johns Hopkins University,
Lawrence Berkeley National Laboratory, Max Planck Institute for Astrophysics, Max
Planck Institute for Extraterrestrial Physics, New Mexico State University, New York
University, Ohio State University, Pennsylvania State University, University of
Portsmouth, Princeton University, the Spanish Participation Group, University of
Tokyo, University of Utah, Vanderbilt University, University of Virginia, University
of Washington, and Yale University.


\begin{thebibliography}{}
\expandafter\ifx\csname natexlab\endcsname\relax\def\natexlab#1{#1}\fi

\bibitem[{{Afonso} {et~al.}(2011){Afonso}, {Greiner}, {Pian}, {Covino},
  {Malesani}, {K{\"u}pc{\"u} Yolda{\c s}}, {Kr{\"u}hler}, {Clemens}, {McBreen},
  {Rau}, {Giannios}, \& {Hjorth}}]{Afonso2011a}
{Afonso}, P., {Greiner}, J., {Pian}, E., {et~al.} 2011, \aap, 526, A154

\bibitem[{{Aihara} {et~al.}(2011){Aihara}, {Allende Prieto}, {An}, {Anderson},
  {Aubourg}, {Balbinot}, {Beers}, {Berlind}, {Bickerton}, {Bizyaev}, {Blanton},
  {Bochanski}, {Bolton}, {Bovy}, {Brandt}, {Brinkmann}, {Brown}, {Brownstein},
  {Busca}, {Campbell}, {Carr}, {Chen}, {Chiappini}, {Comparat}, {Connolly},
  {Cortes}, {Croft}, {Cuesta}, {da Costa}, {Davenport}, {Dawson}, {Dhital},
  {Ealet}, {Ebelke}, {Edmondson}, {Eisenstein}, {Escoffier}, {Esposito},
  {Evans}, {Fan}, {Femen{\'{\i}}a Castell{\'a}}, {Font-Ribera}, {Frinchaboy},
  {Ge}, {Gillespie}, {Gilmore}, {Gonz{\'a}lez Hern{\'a}ndez}, {Gott}, {Gould},
  {Grebel}, {Gunn}, {Hamilton}, {Harding}, {Harris}, {Hawley}, {Hearty}, {Ho},
  {Hogg}, {Holtzman}, {Honscheid}, {Inada}, {Ivans}, {Jiang}, {Johnson},
  {Jordan}, {Jordan}, {Kazin}, {Kirkby}, {Klaene}, {Knapp}, {Kneib},
  {Kochanek}, {Koesterke}, {Kollmeier}, {Kron}, {Lampeitl}, {Lang}, {Le Goff},
  {Lee}, {Lin}, {Long}, {Loomis}, {Lucatello}, {Lundgren}, {Lupton}, {Ma},
  {MacDonald}, {Mahadevan}, {Maia}, {Makler}, {Malanushenko}, {Malanushenko},
  {Mandelbaum}, {Maraston}, {Margala}, {Masters}, {McBride}, {McGehee},
  {McGreer}, {M{\'e}nard}, {Miralda-Escud{\'e}}, {Morrison}, {Mullally},
  {Muna}, {Munn}, {Murayama}, {Myers}, {Naugle}, {Neto}, {Nguyen}, {Nichol},
  {O'Connell}, {Ogando}, {Olmstead}, {Oravetz}, {Padmanabhan},
  {Palanque-Delabrouille}, {Pan}, {Pandey}, {P{\^a}ris}, {Percival},
  {Petitjean}, {Pfaffenberger}, {Pforr}, {Phleps}, {Pichon}, {Pieri}, {Prada},
  {Price-Whelan}, {Raddick}, {Ramos}, {Reyl{\'e}}, {Rich}, {Richards}, {Rix},
  {Robin}, {Rocha-Pinto}, {Rockosi}, {Roe}, {Rollinde}, {Ross}, {Ross},
  {Rossetto}, {S{\'a}nchez}, {Sayres}, {Schlegel}, {Schlesinger}, {Schmidt},
  {Schneider}, {Sheldon}, {Shu}, {Simmerer}, {Simmons}, {Sivarani}, {Snedden},
  {Sobeck}, {Steinmetz}, {Strauss}, {Szalay}, {Tanaka}, {Thakar}, {Thomas},
  {Tinker}, {Tofflemire}, {Tojeiro}, {Tremonti}, {Vandenberg}, {Vargas
  Maga{\~n}a}, {Verde}, {Vogt}, {Wake}, {Wang}, {Weaver}, {Weinberg}, {White},
  {White}, {Yanny}, {Yasuda}, {Yeche}, \& {Zehavi}}]{Aihara2011a}
{Aihara}, H., {Allende Prieto}, C., {An}, D., {et~al.} 2011, \apjs, 193, 29

\bibitem[{{Alavi} {et~al.}(2014){Alavi}, {Siana}, {Richard}, {Stark},
  {Scarlata}, {Teplitz}, {Freeman}, {Dominguez}, {Rafelski}, {Robertson}, \&
  {Kewley}}]{Alavi2014a}
{Alavi}, A., {Siana}, B., {Richard}, J., {et~al.} 2014, \apj, 780, 143

\bibitem[{{Arnouts} {et~al.}(1999){Arnouts}, {Cristiani}, {Moscardini},
  {Matarrese}, {Lucchin}, {Fontana}, \& {Giallongo}}]{Arnouts1999a}
{Arnouts}, S., {Cristiani}, S., {Moscardini}, L., {et~al.} 1999, \mnras, 310,
  540

\bibitem[{{Arnouts} {et~al.}(2005){Arnouts}, {Schiminovich}, {Ilbert},
  {Tresse}, {Milliard}, {Treyer}, {Bardelli}, {Budavari}, {Wyder}, {Zucca}, {Le
  F{\`e}vre}, {Martin}, {Vettolani}, {Adami}, {Arnaboldi}, {Barlow}, {Bianchi},
  {Bolzonella}, {Bottini}, {Byun}, {Cappi}, {Charlot}, {Contini}, {Donas},
  {Forster}, {Foucaud}, {Franzetti}, {Friedman}, {Garilli}, {Gavignaud},
  {Guzzo}, {Heckman}, {Hoopes}, {Iovino}, {Jelinsky}, {Le Brun}, {Lee},
  {Maccagni}, {Madore}, {Malina}, {Marano}, {Marinoni}, {McCracken}, {Mazure},
  {Meneux}, {Merighi}, {Morrissey}, {Neff}, {Paltani}, {Pell{\`o}}, {Picat},
  {Pollo}, {Pozzetti}, {Radovich}, {Rich}, {Scaramella}, {Scodeggio},
  {Seibert}, {Siegmund}, {Small}, {Szalay}, {Welsh}, {Xu}, {Zamorani}, \&
  {Zanichelli}}]{Arnouts2005a}
{Arnouts}, S., {Schiminovich}, D., {Ilbert}, O., {et~al.} 2005, \apj, 619, L43

\bibitem[{{Bertin} \& {Arnouts}(1996)}]{Bertin1996a}
{Bertin}, E., \& {Arnouts}, S. 1996, \aaps, 117, 393

\bibitem[{{Blanton} \& {Roweis}(2007)}]{Blanton2007a}
{Blanton}, M.~R., \& {Roweis}, S. 2007, \aj, 133, 734

\bibitem[{{Bloom} {et~al.}(2002){Bloom}, {Kulkarni}, \&
  {Djorgovski}}]{Bloom2002a}
{Bloom}, J.~S., {Kulkarni}, S.~R., \& {Djorgovski}, S.~G. 2002, \aj, 123, 1111

\bibitem[{{Bouwens} {et~al.}(2007){Bouwens}, {Illingworth}, {Franx}, \&
  {Ford}}]{Bouwens2007a}
{Bouwens}, R.~J., {Illingworth}, G.~D., {Franx}, M., \& {Ford}, H. 2007, \apj,
  670, 928

\bibitem[{{Bouwens} {et~al.}(2009){Bouwens}, {Illingworth}, {Franx}, {Chary},
  {Meurer}, {Conselice}, {Ford}, {Giavalisco}, \& {van Dokkum}}]{Bouwens2009a}
{Bouwens}, R.~J., {Illingworth}, G.~D., {Franx}, M., {et~al.} 2009, \apj, 705,
  936

\bibitem[{{Bouwens} {et~al.}(2010{\natexlab{a}}){Bouwens}, {Illingworth},
  {Oesch}, {Stiavelli}, {van Dokkum}, {Trenti}, {Magee}, {Labb{\'e}}, {Franx},
  {Carollo}, \& {Gonzalez}}]{Bouwens2010a}
{Bouwens}, R.~J., {Illingworth}, G.~D., {Oesch}, P.~A., {et~al.}
  2010{\natexlab{a}}, \apjl, 709, L133

\bibitem[{{Bouwens} {et~al.}(2010{\natexlab{b}}){Bouwens}, {Illingworth},
  {Oesch}, {Trenti}, {Stiavelli}, {Carollo}, {Franx}, {van Dokkum},
  {Labb{\'e}}, \& {Magee}}]{Bouwens2010b}
---. 2010{\natexlab{b}}, \apj, 708, L69

\bibitem[{{Bouwens} {et~al.}(2012){Bouwens}, {Illingworth}, {Oesch}, {Franx},
  {Labb{\'e}}, {Trenti}, {van Dokkum}, {Carollo}, {Gonz{\'a}lez}, {Smit}, \&
  {Magee}}]{Bouwens2012a}
---. 2012, \apj, 754, 83

\bibitem[{{Bouwens} {et~al.}(2014){Bouwens}, {Illingworth}, {Oesch},
  {Labb{\'e}}, {van Dokkum}, {Trenti}, {Franx}, {Smit}, {Gonzalez}, \&
  {Magee}}]{Bouwens2014b}
---. 2014, \apj, 793, 115

\bibitem[{{Breeveld} {et~al.}(2011){Breeveld}, {Landsman}, {Holland}, {Roming},
  {Kuin}, \& {Page}}]{Breeveld2011a}
{Breeveld}, A.~A., {Landsman}, W., {Holland}, S.~T., {et~al.} 2011, in American
  Institute of Physics Conference Series, Vol. 1358, American Institute of
  Physics Conference Series, ed. J.~E. {McEnery}, J.~L. {Racusin}, \&
  N.~{Gehrels}, 373

\bibitem[{{Bruzual} \& {Charlot}(2003)}]{Bruzual2003a}
{Bruzual}, G., \& {Charlot}, S. 2003, \mnras, 344, 1000

\bibitem[{{Butler} {et~al.}(2010){Butler}, {Bloom}, \&
  {Poznanski}}]{Butler2010a}
{Butler}, N.~R., {Bloom}, J.~S., \& {Poznanski}, D. 2010, \apj, 711, 495

\bibitem[{{Butler} {et~al.}(2007){Butler}, {Kocevski}, {Bloom}, \&
  {Curtis}}]{Butler2007a}
{Butler}, N.~R., {Kocevski}, D., {Bloom}, J.~S., \& {Curtis}, J.~L. 2007, \apj,
  671, 656

\bibitem[{{Calzetti} {et~al.}(2000){Calzetti}, {Armus}, {Bohlin}, {Kinney},
  {Koornneef}, \& {Storchi-Bergmann}}]{Calzetti2000a}
{Calzetti}, D., {Armus}, L., {Bohlin}, R.~C., {et~al.} 2000, \apj, 533, 682

\bibitem[{{Cantiello} {et~al.}(2007){Cantiello}, {Yoon}, {Langer}, \&
  {Livio}}]{Cantiello2007a}
{Cantiello}, M., {Yoon}, S.-C., {Langer}, N., \& {Livio}, M. 2007, \aap, 465,
  L29

\bibitem[{{Carilli} \& {Walter}(2013)}]{Carilli2013a}
{Carilli}, C.~L., \& {Walter}, F. 2013, \araa, 51, 105

\bibitem[{{Chen} {et~al.}(2009){Chen}, {Perley}, {Pollack}, {Prochaska},
  {Bloom}, {Dessauges-Zavadsky}, {Pettini}, {Lopez}, {Dall'aglio}, \&
  {Becker}}]{Chen2009a}
{Chen}, H., {Perley}, D.~A., {Pollack}, L.~K., {et~al.} 2009, \apj, 691, 152

\bibitem[{{Chen}(2012)}]{Chen2012a}
{Chen}, H.-W. 2012, \mnras, 419, 3039

\bibitem[{{Christensen} {et~al.}(2004){Christensen}, {Hjorth}, \&
  {Gorosabel}}]{Christensen2004a}
{Christensen}, L., {Hjorth}, J., \& {Gorosabel}, J. 2004, \aap, 425, 913

\bibitem[{{Cucchiara} {et~al.}(2015){Cucchiara}, {Fumagalli}, {Rafelski},
  {Kocevski}, {Prochaska}, {Cooke}, \& {Becker}}]{Cucchiara2015a}
{Cucchiara}, A., {Fumagalli}, M., {Rafelski}, M., {et~al.} 2015, \apj, 804, 51

\bibitem[{{Daddi} {et~al.}(2007){Daddi}, {Dickinson}, {Morrison}, {Chary},
  {Cimatti}, {Elbaz}, {Frayer}, {Renzini}, {Pope}, {Alexander}, {Bauer},
  {Giavalisco}, {Huynh}, {Kurk}, \& {Mignoli}}]{Daddi2007a}
{Daddi}, E., {Dickinson}, M., {Morrison}, G., {et~al.} 2007, \apj, 670, 156

\bibitem[{{Elbaz} {et~al.}(2011){Elbaz}, {Dickinson}, {Hwang},
  {D{\'{\i}}az-Santos}, {Magdis}, {Magnelli}, {Le Borgne}, {Galliano},
  {Pannella}, {Chanial}, {Armus}, {Charmandaris}, {Daddi}, {Aussel}, {Popesso},
  {Kartaltepe}, {Altieri}, {Valtchanov}, {Coia}, {Dannerbauer}, {Dasyra},
  {Leiton}, {Mazzarella}, {Alexander}, {Buat}, {Burgarella}, {Chary}, {Gilli},
  {Ivison}, {Juneau}, {Le Floc'h}, {Lutz}, {Morrison}, {Mullaney}, {Murphy},
  {Pope}, {Scott}, {Brodwin}, {Calzetti}, {Cesarsky}, {Charlot}, {Dole},
  {Eisenhardt}, {Ferguson}, {F{\"o}rster Schreiber}, {Frayer}, {Giavalisco},
  {Huynh}, {Koekemoer}, {Papovich}, {Reddy}, {Surace}, {Teplitz}, {Yun}, \&
  {Wilson}}]{Elbaz2011a}
{Elbaz}, D., {Dickinson}, M., {Hwang}, H.~S., {et~al.} 2011, \aap, 533, A119

\bibitem[{{Elliott} {et~al.}(2012){Elliott}, {Greiner}, {Khochfar}, {Schady},
  {Johnson}, \& {Rau}}]{Elliott2012a}
{Elliott}, J., {Greiner}, J., {Khochfar}, S., {et~al.} 2012, \aap, 539, A113

\bibitem[{{Elliott} {et~al.}(2013){Elliott}, {Kr{\"u}hler}, {Greiner},
  {Savaglio}, {Olivares}, {Rau}, {de Ugarte Postigo},
  {S{\'a}nchez-Ram{\'{\i}}rez}, {Wiersema}, {Schady}, {Kann}, {Filgas},
  {Nardini}, {Berger}, {Fox}, {Gorosabel}, {Klose}, {Levan}, {Nicuesa
  Guelbenzu}, {Rossi}, {Schmidl}, {Sudilovsky}, {Tanvir}, \&
  {Th{\"o}ne}}]{Elliott2013a}
{Elliott}, J., {Kr{\"u}hler}, T., {Greiner}, J., {et~al.} 2013, \aap, 556, A23

\bibitem[{{Erb} {et~al.}(2006){Erb}, {Shapley}, {Pettini}, {Steidel}, {Reddy},
  \& {Adelberger}}]{Erb2006a}
{Erb}, D.~K., {Shapley}, A.~E., {Pettini}, M., {et~al.} 2006, \apj, 644, 813

\bibitem[{{Finkelstein} {et~al.}(2012){Finkelstein}, {Papovich}, {Salmon},
  {Finlator}, {Dickinson}, {Ferguson}, {Giavalisco}, {Koekemoer}, {Reddy},
  {Bassett}, {Conselice}, {Dunlop}, {Faber}, {Grogin}, {Hathi}, {Kocevski},
  {Lai}, {Lee}, {McLure}, {Mobasher}, \& {Newman}}]{Finkelstein2012a}
{Finkelstein}, S.~L., {Papovich}, C., {Salmon}, B., {et~al.} 2012, \apj, 756,
  164

\bibitem[{{Foster} {et~al.}(2012){Foster}, {Hopkins}, {Gunawardhana},
  {Lara-L{\'o}pez}, {Sharp}, {Steele}, {Taylor}, {Driver}, {Baldry}, {Bamford},
  {Liske}, {Loveday}, {Norberg}, {Peacock}, {Alpaslan}, {Bauer},
  {Bland-Hawthorn}, {Brough}, {Cameron}, {Colless}, {Conselice}, {Croom},
  {Frenk}, {Hill}, {Jones}, {Kelvin}, {Kuijken}, {Nichol}, {Owers},
  {Parkinson}, {Pimbblet}, {Popescu}, {Prescott}, {Robotham}, {Lopez-Sanchez},
  {Sutherland}, {Thomas}, {Tuffs}, {van Kampen}, \& {Wijesinghe}}]{Foster2012a}
{Foster}, C., {Hopkins}, A.~M., {Gunawardhana}, M., {et~al.} 2012, \aap, 547,
  A79

\bibitem[{{Fruchter} {et~al.}(2006){Fruchter}, {Levan}, {Strolger},
  {Vreeswijk}, {Thorsett}, {Bersier}, {Burud}, {Castro Cer{\'o}n},
  {Castro-Tirado}, {Conselice}, {Dahlen}, {Ferguson}, {Fynbo}, {Garnavich},
  {Gibbons}, {Gorosabel}, {Gull}, {Hjorth}, {Holland}, {Kouveliotou}, {Levay},
  {Livio}, {Metzger}, {Nugent}, {Petro}, {Pian}, {Rhoads}, {Riess}, {Sahu},
  {Smette}, {Tanvir}, {Wijers}, \& {Woosley}}]{Fruchter2006a}
{Fruchter}, A.~S., {Levan}, A.~J., {Strolger}, L., {et~al.} 2006, \nat, 441,
  463

\bibitem[{{Fynbo} {et~al.}(2002){Fynbo}, {M{\"o}ller}, {Thomsen}, {Hjorth},
  {Gorosabel}, {Andersen}, {Egholm}, {Holland}, {Jensen}, {Pedersen}, \&
  {Weidinger}}]{Fynbo2002a}
{Fynbo}, J.~P.~U., {M{\"o}ller}, P., {Thomsen}, B., {et~al.} 2002, \aap, 388,
  425

\bibitem[{{Fynbo} {et~al.}(2009){Fynbo}, {Jakobsson}, {Prochaska}, {Malesani},
  {Ledoux}, {de Ugarte Postigo}, {Nardini}, {Vreeswijk}, {Wiersema}, {Hjorth},
  {Sollerman}, {Chen}, {Th{\"o}ne}, {Bj{\"o}rnsson}, {Bloom}, {Castro-Tirado},
  {Christensen}, {De Cia}, {Fruchter}, {Gorosabel}, {Graham}, {Jaunsen},
  {Jensen}, {Kann}, {Kouveliotou}, {Levan}, {Maund}, {Masetti},
  {Milvang-Jensen}, {Palazzi}, {Perley}, {Pian}, {Rol}, {Schady}, {Starling},
  {Tanvir}, {Watson}, {Xu}, {Augusteijn}, {Grundahl}, {Telting}, \&
  {Quirion}}]{Fynbo2009a}
{Fynbo}, J.~P.~U., {Jakobsson}, P., {Prochaska}, J.~X., {et~al.} 2009, \apjs,
  185, 526

\bibitem[{{Fynbo} {et~al.}(1999){Fynbo}, {M{\o}ller}, \& {Warren}}]{Fynbo1999a}
{Fynbo}, J.~U., {M{\o}ller}, P., \& {Warren}, S.~J. 1999, \mnras, 305, 849

\bibitem[{{Graham} \& {Fruchter}(2013)}]{Graham2013a}
{Graham}, J.~F., \& {Fruchter}, A.~S. 2013, \apj, 774, 119

\bibitem[{{Greiner} {et~al.}(2008){Greiner}, {Bornemann}, {Clemens}, {Deuter},
  {Hasinger}, {Honsberg}, {Huber}, {Huber}, {Krauss}, {Kr{\"u}hler},
  {K{\"u}pc{\"u} Yolda{\c s}}, {Mayer-Hasselwander}, {Mican}, {Primak},
  {Schrey}, {Steiner}, {Szokoly}, {Th{\"o}ne}, {Yolda{\c s}}, {Klose}, {Laux},
  \& {Winkler}}]{Greiner2008a}
{Greiner}, J., {Bornemann}, W., {Clemens}, C., {et~al.} 2008, \pasp, 120, 405

\bibitem[{{Greiner} {et~al.}(2015){Greiner}, {Fox}, {Schady}, {Kr{\"u}hler},
  {Trenti}, {Cikota}, {Bolmer}, {Elliott}, {Delvaux}, {Perna}, {Afonso},
  {Kann}, {Klose}, {Savaglio}, {Schmidl}, {Schweyer}, {Tanga}, \&
  {Varela}}]{Greiner2015a}
{Greiner}, J., {Fox}, D.~B., {Schady}, P., {et~al.} 2015, \apj, submitted,
  arXiv:1503.05323

\bibitem[{{Grupe} {et~al.}(2007){Grupe}, {Nousek}, {vanden Berk}, {Roming},
  {Burrows}, {Godet}, {Osborne}, \& {Gehrels}}]{Grupe2007a}
{Grupe}, D., {Nousek}, J.~A., {vanden Berk}, D.~E., {et~al.} 2007, \aj, 133,
  2216

\bibitem[{{Haehnelt} {et~al.}(2000){Haehnelt}, {Steinmetz}, \&
  {Rauch}}]{Haehnelt2000a}
{Haehnelt}, M.~G., {Steinmetz}, M., \& {Rauch}, M. 2000, \apj, 534, 594

\bibitem[{{Hathi} {et~al.}(2008){Hathi}, {Malhotra}, \& {Rhoads}}]{Hathi2008a}
{Hathi}, N.~P., {Malhotra}, S., \& {Rhoads}, J.~E. 2008, \apj, 673, 686

\bibitem[{{Hathi} {et~al.}(2013){Hathi}, {Cohen}, {Ryan}, {Finkelstein},
  {McCarthy}, {Windhorst}, {Yan}, {Koekemoer}, {Rutkowski}, {O'Connell},
  {Straughn}, {Balick}, {Bond}, {Calzetti}, {Disney}, {Dopita}, {Frogel},
  {Hall}, {Holtzman}, {Kimble}, {Paresce}, {Saha}, {Silk}, {Trauger}, {Walker},
  {Whitmore}, \& {Young}}]{Hathi2013a}
{Hathi}, N.~P., {Cohen}, S.~H., {Ryan}, Jr., R.~E., {et~al.} 2013, \apj, 765,
  88

\bibitem[{{Hjorth} {et~al.}(2012){Hjorth}, {Malesani}, {Jakobsson}, {Jaunsen},
  {Fynbo}, {Gorosabel}, {Kr{\"u}hler}, {Levan}, {Micha{\l}owski},
  {Milvang-Jensen}, {M{\o}ller}, {Schulze}, {Tanvir}, \&
  {Watson}}]{Hjorth2012a}
{Hjorth}, J., {Malesani}, D., {Jakobsson}, P., {et~al.} 2012, \apj, 756, 187

\bibitem[{{Hogg} {et~al.}(1997){Hogg}, {Pahre}, {McCarthy}, {Cohen},
  {Blandford}, {Smail}, \& {Soifer}}]{Hogg1997a}
{Hogg}, D.~W., {Pahre}, M.~A., {McCarthy}, J.~K., {et~al.} 1997, \mnras, 288,
  404

\bibitem[{{Hopkins}(2004)}]{Hopkins2004a}
{Hopkins}, A.~M. 2004, \apj, 615, 209

\bibitem[{{Hopkins} \& {Beacom}(2006)}]{Hopkins2006a}
{Hopkins}, A.~M., \& {Beacom}, J.~F. 2006, \apj, 651, 142

\bibitem[{{Hunt} {et~al.}(2014){Hunt}, {Palazzi}, {Micha{\l}owski}, {Rossi},
  {Savaglio}, {Basa}, {Berta}, {Bianchi}, {Covino}, {D'Elia}, {Ferrero},
  {G{\"o}tz}, {Greiner}, {Klose}, {Le Borgne}, {Le Floc'h}, {Pian},
  {Piranomonte}, {Schady}, \& {Vergani}}]{Hunt2014a}
{Hunt}, L.~K., {Palazzi}, E., {Micha{\l}owski}, M.~J., {et~al.} 2014, \aap,
  565, A112

\bibitem[{{Ilbert} {et~al.}(2006){Ilbert}, {Arnouts}, {McCracken},
  {Bolzonella}, {Bertin}, {Le F{\`e}vre}, {Mellier}, {Zamorani}, {Pell{\`o}},
  {Iovino}, {Tresse}, {Le Brun}, {Bottini}, {Garilli}, {Maccagni}, {Picat},
  {Scaramella}, {Scodeggio}, {Vettolani}, {Zanichelli}, {Adami}, {Bardelli},
  {Cappi}, {Charlot}, {Ciliegi}, {Contini}, {Cucciati}, {Foucaud}, {Franzetti},
  {Gavignaud}, {Guzzo}, {Marano}, {Marinoni}, {Mazure}, {Meneux}, {Merighi},
  {Paltani}, {Pollo}, {Pozzetti}, {Radovich}, {Zucca}, {Bondi}, {Bongiorno},
  {Busarello}, {de La Torre}, {Gregorini}, {Lamareille}, {Mathez}, {Merluzzi},
  {Ripepi}, {Rizzo}, \& {Vergani}}]{Ilbert2006a}
{Ilbert}, O., {Arnouts}, S., {McCracken}, H.~J., {et~al.} 2006, \aap, 457, 841

\bibitem[{{Jakobsson} {et~al.}(2005){Jakobsson}, {Bj{\"o}rnsson}, {Fynbo},
  {J{\'o}hannesson}, {Hjorth}, {Thomsen}, {M{\o}ller}, {Watson}, {Jensen},
  {{\"O}stlin}, {Gorosabel}, \& {Gudmundsson}}]{Jakobsson2005a}
{Jakobsson}, P., {Bj{\"o}rnsson}, G., {Fynbo}, J.~P.~U., {et~al.} 2005, \mnras,
  362, 245

\bibitem[{{Jakobsson} {et~al.}(2012){Jakobsson}, {Hjorth}, {Malesani},
  {Chapman}, {Fynbo}, {Tanvir}, {Milvang-Jensen}, {Vreeswijk}, {Letawe}, \&
  {Starling}}]{Jakobsson2012a}
{Jakobsson}, P., {Hjorth}, J., {Malesani}, D., {et~al.} 2012, \apj, 752, 62

\bibitem[{{Kistler} {et~al.}(2009){Kistler}, {Y{\"u}ksel}, {Beacom}, {Hopkins},
  \& {Wyithe}}]{Kistler2009a}
{Kistler}, M.~D., {Y{\"u}ksel}, H., {Beacom}, J.~F., {Hopkins}, A.~M., \&
  {Wyithe}, J.~S.~B. 2009, \apjl, 705, L104

\bibitem[{{Kistler} {et~al.}(2008){Kistler}, {Y{\"u}ksel}, {Beacom}, \&
  {Stanek}}]{Kistler2008a}
{Kistler}, M.~D., {Y{\"u}ksel}, H., {Beacom}, J.~F., \& {Stanek}, K.~Z. 2008,
  \apjl, 673, L119

\bibitem[{{Kocevski} \& {West}(2011)}]{Kocevski2011a}
{Kocevski}, D., \& {West}, A.~A. 2011, \apj, 735, L8

\bibitem[{{Koekemoer} {et~al.}(2003){Koekemoer}, {Fruchter}, {Hook}, \&
  {Hack}}]{Koekemoer2003a}
{Koekemoer}, A.~M., {Fruchter}, A.~S., {Hook}, R.~N., \& {Hack}, W. 2003, in
  HST Calibration Workshop : Hubble after the Installation of the ACS and the
  NICMOS Cooling System, ed. S.~{Arribas}, A.~{Koekemoer}, \& B.~{Whitmore},
  337

\bibitem[{{Kohn} {et~al.}(2015){Kohn}, {Micha{\l}owski}, {Bourne}, {Baes},
  {Fritz}, {Cooray}, {De Looze}, {De Zotti}, {Dannerbauer}, {Dunne}, {Dye},
  {Eales}, {Furlanetto}, {Gonzalez-Nuevo}, {Ibar}, {Ivison}, {Maddox}, {Scott},
  {Smith}, {Smith}, {Symeonidis}, \& {Valiante}}]{Kohn2015a}
{Kohn}, S.~A., {Micha{\l}owski}, M.~J., {Bourne}, N., {et~al.} 2015, \mnras,
  448, 1494

\bibitem[{{Kr{\"u}hler} {et~al.}(2008){Kr{\"u}hler}, {K{\"u}pc{\"u} Yolda{\c
  s}}, {Greiner}, {Clemens}, {McBreen}, {Primak}, {Savaglio}, {Yolda{\c s}},
  {Szokoly}, \& {Klose}}]{Kruehler2008a}
{Kr{\"u}hler}, T., {K{\"u}pc{\"u} Yolda{\c s}}, A., {Greiner}, J., {et~al.}
  2008, \apj, 685, 376

\bibitem[{{Kr{\"u}hler} {et~al.}(2011){Kr{\"u}hler}, {Greiner}, {Schady},
  {Savaglio}, {Afonso}, {Clemens}, {Elliott}, {Filgas}, {Gruber}, {Kann},
  {Klose}, {K{\"u}pc{\"u}-Yolda{\c s}}, {McBreen}, {Olivares}, {Pierini},
  {Rau}, {Rossi}, {Nardini}, {Nicuesa Guelbenzu}, {Sudilovsky}, \&
  {Updike}}]{Kruehler2011a}
{Kr{\"u}hler}, T., {Greiner}, J., {Schady}, P., {et~al.} 2011, \aap, 534, A108

\bibitem[{{Kr{\"u}hler} {et~al.}(2012){Kr{\"u}hler}, {Malesani},
  {Milvang-Jensen}, {Fynbo}, {Hjorth}, {Jakobsson}, {Levan}, {Sparre},
  {Tanvir}, \& {Watson}}]{Kruehler2012a}
{Kr{\"u}hler}, T., {Malesani}, D., {Milvang-Jensen}, B., {et~al.} 2012, \apj,
  758, 46

\bibitem[{{Kr{\"u}hler} {et~al.}(2015){Kr{\"u}hler}, {Malesani}, {Fynbo},
  {Hartoog}, {Hjorth}, {Jakobsson}, {Perley}, {Rossi}, {Schady}, {Schulze},
  {Tanvir}, {Vergani}, {Wiersema}, {Afonso}, {Bolmer}, {Cano}, {Covino},
  {D'Elia}, {de Ugarte Postigo}, {Filgas}, {Friis}, {Graham}, {Greiner},
  {Goldoni}, {Gomboc}, {Hammer}, {Japelj}, {Kann}, {Kaper}, {Klose}, {Levan},
  {Leloudas}, {Milvang-Jensen}, {Nicuesa Guelbenzu}, {Palazzi}, {Pian},
  {Piranomonte}, {Sanchez-Ramirez}, {Savaglio}, {Selsing}, {Tagliaferri},
  {Vreeswijk}, {Watson}, \& {Xu}}]{Kruehler2015a}
{Kr{\"u}hler}, T., {Malesani}, D., {Fynbo}, J.~P.~U., {et~al.} 2015, \aap,
  accepted, arXiv:1505.06743

\bibitem[{{Larson} {et~al.}(2011){Larson}, {Dunkley}, {Hinshaw}, {Komatsu},
  {Nolta}, {Bennett}, {Gold}, {Halpern}, {Hill}, {Jarosik}, {Kogut}, {Limon},
  {Meyer}, {Odegard}, {Page}, {Smith}, {Spergel}, {Tucker}, {Weiland},
  {Wollack}, \& {Wright}}]{Larson2011a}
{Larson}, D., {Dunkley}, J., {Hinshaw}, G., {et~al.} 2011, \apjs, 192, 16

\bibitem[{{Le Floc'h} {et~al.}(2003){Le Floc'h}, {Duc}, {Mirabel}, {Sanders},
  {Bosch}, {Diaz}, {Donzelli}, {Rodrigues}, {Courvoisier}, {Greiner},
  {Mereghetti}, {Melnick}, {Maza}, \& {Minniti}}]{LeFloch2003a}
{Le Floc'h}, E., {Duc}, P.-A., {Mirabel}, I.~F., {et~al.} 2003, \aap, 400, 499

\bibitem[{{Lee} {et~al.}(2006){Lee}, {Skillman}, {Cannon}, {Jackson}, {Gehrz},
  {Polomski}, \& {Woodward}}]{Lee2006a}
{Lee}, H., {Skillman}, E.~D., {Cannon}, J.~M., {et~al.} 2006, \apj, 647, 970

\bibitem[{{Levesque} {et~al.}(2010){Levesque}, {Kewley}, {Graham}, \&
  {Fruchter}}]{Levesque2010a}
{Levesque}, E.~M., {Kewley}, L.~J., {Graham}, J.~F., \& {Fruchter}, A.~S. 2010,
  \apjl, 712, L26

\bibitem[{{Magdis} {et~al.}(2010){Magdis}, {Rigopoulou}, {Huang}, \&
  {Fazio}}]{Magdis2010a}
{Magdis}, G.~E., {Rigopoulou}, D., {Huang}, J.-S., \& {Fazio}, G.~G. 2010,
  \mnras, 401, 1521

\bibitem[{{Mangano} {et~al.}(2007){Mangano}, {Holland}, {Malesani}, {Troja},
  {Chincarini}, {Zhang}, {La Parola}, {Brown}, {Burrows}, {Campana}, {Capalbi},
  {Cusumano}, {Della Valle}, {Gehrels}, {Giommi}, {Grupe}, {Guidorzi}, {Mineo},
  {Moretti}, {Osborne}, {Pandey}, {Perri}, {Romano}, {Roming}, \&
  {Tagliaferri}}]{Mangano2007a}
{Mangano}, V., {Holland}, S.~T., {Malesani}, D., {et~al.} 2007, \aap, 470, 105

\bibitem[{{Meurer} {et~al.}(1999){Meurer}, {Heckman}, \&
  {Calzetti}}]{Meurer1999a}
{Meurer}, G.~R., {Heckman}, T.~M., \& {Calzetti}, D. 1999, \apj, 521, 64

\bibitem[{{Micha{\l}owski} {et~al.}(2008){Micha{\l}owski}, {Hjorth}, {Castro
  Cer{\'o}n}, \& {Watson}}]{Michalowski2008a}
{Micha{\l}owski}, M.~J., {Hjorth}, J., {Castro Cer{\'o}n}, J.~M., \& {Watson},
  D. 2008, \apj, 672, 817

\bibitem[{{Micha{\l}owski} {et~al.}(2012){Micha{\l}owski}, {Kamble}, {Hjorth},
  {Malesani}, {Reinfrank}, {Bonavera}, {Castro Cer{\'o}n}, {Ibar}, {Dunlop},
  {Fynbo}, {Garrett}, {Jakobsson}, {Kaplan}, {Kr{\"u}hler}, {Levan},
  {Massardi}, {Pal}, {Sollerman}, {Tanvir}, {van der Horst}, {Watson}, \&
  {Wiersema}}]{Michalowski2012a}
{Micha{\l}owski}, M.~J., {Kamble}, A., {Hjorth}, J., {et~al.} 2012, \apj, 755,
  85

\bibitem[{{Milvang-Jensen} {et~al.}(2012){Milvang-Jensen}, {Fynbo}, {Malesani},
  {Hjorth}, {Jakobsson}, \& {M{\o}ller}}]{MilvangJensen2012a}
{Milvang-Jensen}, B., {Fynbo}, J.~P.~U., {Malesani}, D., {et~al.} 2012, \apj,
  756, 25

\bibitem[{{Modjaz} {et~al.}(2008){Modjaz}, {Kewley}, {Kirshner}, {Stanek},
  {Challis}, {Garnavich}, {Greene}, {Kelly}, \& {Prieto}}]{Modjaz2008a}
{Modjaz}, M., {Kewley}, L., {Kirshner}, R.~P., {et~al.} 2008, \aj, 135, 1136

\bibitem[{{Noeske} {et~al.}(2007){Noeske}, {Weiner}, {Faber}, {Papovich},
  {Koo}, {Somerville}, {Bundy}, {Conselice}, {Newman}, {Schiminovich}, {Le
  Floc'h}, {Coil}, {Rieke}, {Lotz}, {Primack}, {Barmby}, {Cooper}, {Davis},
  {Ellis}, {Fazio}, {Guhathakurta}, {Huang}, {Kassin}, {Martin}, {Phillips},
  {Rich}, {Small}, {Willmer}, \& {Wilson}}]{Noeske2007a}
{Noeske}, K.~G., {Weiner}, B.~J., {Faber}, S.~M., {et~al.} 2007, \apjl, 660,
  L43

\bibitem[{{Oesch} {et~al.}(2010){Oesch}, {Bouwens}, {Carollo}, {Illingworth},
  {Magee}, {Trenti}, {Stiavelli}, {Franx}, {Labb{\'e}}, \& {van
  Dokkum}}]{Oesch2010a}
{Oesch}, P.~A., {Bouwens}, R.~J., {Carollo}, C.~M., {et~al.} 2010, \apjl, 725,
  L150

\bibitem[{{Perley} {et~al.}(2009){Perley}, {Cenko}, {Bloom}, {Chen}, {Butler},
  {Kocevski}, {Prochaska}, {Brodwin}, {Glazebrook}, {Kasliwal}, {Kulkarni},
  {Lopez}, {Ofek}, {Pettini}, {Soderberg}, \& {Starr}}]{Perley2009a}
{Perley}, D.~A., {Cenko}, S.~B., {Bloom}, J.~S., {et~al.} 2009, \aj, 138, 1690

\bibitem[{{Perley} {et~al.}(2013){Perley}, {Levan}, {Tanvir}, {Cenko}, {Bloom},
  {Hjorth}, {Kr{\"u}hler}, {Filippenko}, {Fruchter}, {Fynbo}, {Jakobsson},
  {Kalirai}, {Milvang-Jensen}, {Morgan}, {Prochaska}, \&
  {Silverman}}]{Perley2013a}
{Perley}, D.~A., {Levan}, A.~J., {Tanvir}, N.~R., {et~al.} 2013, \apj, 778, 128

\bibitem[{{Perley} {et~al.}(2015{\natexlab{a}}){Perley}, {Kr{\"u}hler},
  {Schulze}, {de Ugarte Postigo}, {Hjorth}, {Berger}, {Cenko}, {Chary},
  {Cucchiara}, {Ellis}, {Fong}, {Fynbo}, {Gorosabel}, {Greiner}, {Jakobsson},
  {Kim}, {Laskar}, {Levan}, {Micha{\l}owski}, {Milvang-Jensen}, {Tanvir},
  {Th{\"o}ne}, \& {Wiersema}}]{Perley2015b}
{Perley}, D.~A., {Kr{\"u}hler}, T., {Schulze}, S., {et~al.} 2015{\natexlab{a}},
  \apj, submitted, arXiv:1504.02482

\bibitem[{{Perley} {et~al.}(2015{\natexlab{b}}){Perley}, {Tanvir}, {Hjorth},
  {Laskar}, {Berger}, {Chary}, {de Ugarte Postigo}, {Fynbo}, {Kr{\"u}hler},
  {Levan}, {Micha{\l}owski}, \& {Schulze}}]{Perley2015a}
{Perley}, D.~A., {Tanvir}, N.~R., {Hjorth}, J., {et~al.} 2015{\natexlab{b}},
  \apj, submitted, arXiv:1504.02479

\bibitem[{{Pettini} {et~al.}(2002){Pettini}, {Rix}, {Steidel}, {Adelberger},
  {Hunt}, \& {Shapley}}]{Pettini2002a}
{Pettini}, M., {Rix}, S.~A., {Steidel}, C.~C., {et~al.} 2002, \apj, 569, 742

\bibitem[{{Pettini} {et~al.}(2001){Pettini}, {Shapley}, {Steidel}, {Cuby},
  {Dickinson}, {Moorwood}, {Adelberger}, \& {Giavalisco}}]{Pettini2001a}
{Pettini}, M., {Shapley}, A.~E., {Steidel}, C.~C., {et~al.} 2001, \apj, 554,
  981

\bibitem[{{Reddy} \& {Steidel}(2009)}]{Reddy2009a}
{Reddy}, N.~A., \& {Steidel}, C.~C. 2009, \apj, 692, 778

\bibitem[{{Robertson} \& {Ellis}(2012)}]{Robertson2012a}
{Robertson}, B.~E., \& {Ellis}, R.~S. 2012, \apj, 744, 95

\bibitem[{{Rodighiero} {et~al.}(2010){Rodighiero}, {Cimatti}, {Gruppioni},
  {Popesso}, {Andreani}, {Altieri}, {Aussel}, {Berta}, {Bongiovanni},
  {Brisbin}, {Cava}, {Cepa}, {Daddi}, {Dominguez-Sanchez}, {Elbaz}, {Fontana},
  {F{\"o}rster Schreiber}, {Franceschini}, {Genzel}, {Grazian}, {Lutz},
  {Magdis}, {Magliocchetti}, {Magnelli}, {Maiolino}, {Mancini}, {Nordon},
  {Perez Garcia}, {Poglitsch}, {Santini}, {Sanchez-Portal}, {Pozzi},
  {Riguccini}, {Saintonge}, {Shao}, {Sturm}, {Tacconi}, {Valtchanov},
  {Wetzstein}, \& {Wieprecht}}]{Rodighiero2010a}
{Rodighiero}, G., {Cimatti}, A., {Gruppioni}, C., {et~al.} 2010, \aap, 518, L25

\bibitem[{{Rossi} {et~al.}(2012){Rossi}, {Klose}, {Ferrero}, {Greiner},
  {Arnold}, {Gonsalves}, {Hartmann}, {Updike}, {Kann}, {Kr{\"u}hler},
  {Palazzi}, {Savaglio}, {Schulze}, {Afonso}, {Amati}, {Castro-Tirado},
  {Clemens}, {Filgas}, {Gorosabel}, {Hunt}, {K{\"u}pc{\"u} Yolda{\c s}},
  {Masetti}, {Nardini}, {Nicuesa Guelbenzu}, {Olivares}, {Pian}, {Rau},
  {Schady}, {Schmidl}, {Yolda{\c s}}, \& {de Ugarte Postigo}}]{Rossi2012a}
{Rossi}, A., {Klose}, S., {Ferrero}, P., {et~al.} 2012, \aap, 545, A77

\bibitem[{{Sakamoto} {et~al.}(2008){Sakamoto}, {Barthelmy}, {Barbier},
  {Cummings}, {Fenimore}, {Gehrels}, {Hullinger}, {Krimm}, {Markwardt},
  {Palmer}, {Parsons}, {Sato}, {Stamatikos}, {Tueller}, {Ukwatta}, \&
  {Zhang}}]{Sakamoto2008a}
{Sakamoto}, T., {Barthelmy}, S.~D., {Barbier}, L., {et~al.} 2008, \apjs, 175,
  179

\bibitem[{{Salvaterra} {et~al.}(2009){Salvaterra}, {Della Valle}, {Campana},
  {Chincarini}, {Covino}, {D'Avanzo}, {Fern{\'a}ndez-Soto}, {Guidorzi},
  {Mannucci}, {Margutti}, {Th{\"o}ne}, {Antonelli}, {Barthelmy}, {de Pasquale},
  {D'Elia}, {Fiore}, {Fugazza}, {Hunt}, {Maiorano}, {Marinoni}, {Marshall},
  {Molinari}, {Nousek}, {Pian}, {Racusin}, {Stella}, {Amati}, {Andreuzzi},
  {Cusumano}, {Fenimore}, {Ferrero}, {Giommi}, {Guetta}, {Holland}, {Hurley},
  {Israel}, {Mao}, {Markwardt}, {Masetti}, {Pagani}, {Palazzi}, {Palmer},
  {Piranomonte}, {Tagliaferri}, \& {Testa}}]{Salvaterra2009a}
{Salvaterra}, R., {Della Valle}, M., {Campana}, S., {et~al.} 2009, \nat, 461,
  1258

\bibitem[{{Salvaterra} {et~al.}(2012){Salvaterra}, {Campana}, {Vergani},
  {Covino}, {D'Avanzo}, {Fugazza}, {Ghirlanda}, {Ghisellini}, {Melandri},
  {Nava}, {Sbarufatti}, {Flores}, {Piranomonte}, \&
  {Tagliaferri}}]{Salvaterra2012a}
{Salvaterra}, R., {Campana}, S., {Vergani}, S.~D., {et~al.} 2012, \apj, 749, 68

\bibitem[{{Savaglio}(2006)}]{Savaglio2006a}
{Savaglio}, S. 2006, New Journal of Physics, 8, 195

\bibitem[{{Savaglio} {et~al.}(2009){Savaglio}, {Glazebrook}, \& {Le
  Borgne}}]{Savaglio2009a}
{Savaglio}, S., {Glazebrook}, K., \& {Le Borgne}, D. 2009, \apj, 691, 182

\bibitem[{{Savaglio} {et~al.}(2005){Savaglio}, {Glazebrook}, {Le Borgne},
  {Juneau}, {Abraham}, {Chen}, {Crampton}, {McCarthy}, {Carlberg}, {Marzke},
  {Roth}, {J{\o}rgensen}, \& {Murowinski}}]{Savaglio2005a}
{Savaglio}, S., {Glazebrook}, K., {Le Borgne}, D., {et~al.} 2005, \apj, 635,
  260

\bibitem[{{Schady} {et~al.}(2014){Schady}, {Savaglio}, {M{\"u}ller},
  {Kr{\"u}hler}, {Dwelly}, {Palazzi}, {Hunt}, {Greiner}, {Linz},
  {Micha{\l}owski}, {Pierini}, {Piranomonte}, {Vergani}, \&
  {Gear}}]{Schady2014a}
{Schady}, P., {Savaglio}, S., {M{\"u}ller}, T., {et~al.} 2014, \aap, 570, A52

\bibitem[{{Schaerer} {et~al.}(2013){Schaerer}, {de Barros}, \&
  {Sklias}}]{Schaerer2013a}
{Schaerer}, D., {de Barros}, S., \& {Sklias}, P. 2013, \aap, 549, A4

\bibitem[{{Schiminovich} {et~al.}(2005){Schiminovich}, {Ilbert}, {Arnouts},
  {Milliard}, {Tresse}, {Le F{\`e}vre}, {Treyer}, {Wyder}, {Budav{\'a}ri},
  {Zucca}, {Zamorani}, {Martin}, {Adami}, {Arnaboldi}, {Bardelli}, {Barlow},
  {Bianchi}, {Bolzonella}, {Bottini}, {Byun}, {Cappi}, {Contini}, {Charlot},
  {Donas}, {Forster}, {Foucaud}, {Franzetti}, {Friedman}, {Garilli},
  {Gavignaud}, {Guzzo}, {Heckman}, {Hoopes}, {Iovino}, {Jelinsky}, {Le Brun},
  {Lee}, {Maccagni}, {Madore}, {Malina}, {Marano}, {Marinoni}, {McCracken},
  {Mazure}, {Meneux}, {Morrissey}, {Neff}, {Paltani}, {Pell{\`o}}, {Picat},
  {Pollo}, {Pozzetti}, {Radovich}, {Rich}, {Scaramella}, {Scodeggio},
  {Seibert}, {Siegmund}, {Small}, {Szalay}, {Vettolani}, {Welsh}, {Xu}, \&
  {Zanichelli}}]{Schiminovich2005a}
{Schiminovich}, D., {Ilbert}, O., {Arnouts}, S., {et~al.} 2005, \apjl, 619, L47

\bibitem[{{Schlegel} {et~al.}(1998){Schlegel}, {Finkbeiner}, \&
  {Davis}}]{Schlegel1998a}
{Schlegel}, D.~J., {Finkbeiner}, D.~P., \& {Davis}, M. 1998, \apj, 500, 525

\bibitem[{{Schmidt}(2009)}]{Schmidt2009a}
{Schmidt}, M. 2009, \apj, 700, 633

\bibitem[{{Schulze} {et~al.}(2012){Schulze}, {Fynbo}, {Milvang-Jensen},
  {Rossi}, {Jakobsson}, {Ledoux}, {De Cia}, {Kr{\"u}hler}, {Mehner},
  {Bj{\"o}rnsson}, {Chen}, {Vreeswijk}, {Perley}, {Hjorth}, {Levan}, {Tanvir},
  {Ellison}, {M{\o}ller}, {Worseck}, {Chapman}, {Dall'Aglio}, \&
  {Letawe}}]{Schulze2012a}
{Schulze}, S., {Fynbo}, J.~P.~U., {Milvang-Jensen}, B., {et~al.} 2012, \aap,
  546, A20

\bibitem[{{Schulze} {et~al.}(2014){Schulze}, {Malesani}, {Cucchiara}, {Tanvir},
  {Kr{\"u}hler}, {de Ugarte Postigo}, {Leloudas}, {Lyman}, {Bersier},
  {Wiersema}, {Perley}, {Schady}, {Gorosabel}, {Anderson}, {Castro-Tirado},
  {Cenko}, {De Cia}, {Ellerbroek}, {Fynbo}, {Greiner}, {Hjorth}, {Kann},
  {Kaper}, {Klose}, {Levan}, {Mart{\'{\i}}n}, {O'Brien}, {Page}, {Pignata},
  {Rapaport}, {S{\'a}nchez-Ram{\'{\i}}rez}, {Sollerman}, {Smith}, {Sparre},
  {Th{\"o}ne}, {Watson}, {Xu}, {Bauer}, {Bayliss}, {Bj{\"o}rnsson}, {Bremer},
  {Cano}, {Covino}, {D'Elia}, {Frail}, {Geier}, {Goldoni}, {Hartoog},
  {Jakobsson}, {Korhonen}, {Lee}, {Milvang-Jensen}, {Nardini}, {Nicuesa
  Guelbenzu}, {Oguri}, {Pandey}, {Petitpas}, {Rossi}, {Sandberg}, {Schmidl},
  {Tagliaferri}, {Tilanus}, {Winters}, {Wright}, \& {Wuyts}}]{Schulze2014a}
{Schulze}, S., {Malesani}, D., {Cucchiara}, A., {et~al.} 2014, \aap, 566, A102

\bibitem[{{Scoville} {et~al.}(2007){Scoville}, {Aussel}, {Brusa}, {Capak},
  {Carollo}, {Elvis}, {Giavalisco}, {Guzzo}, {Hasinger}, {Impey}, {Kneib},
  {LeFevre}, {Lilly}, {Mobasher}, {Renzini}, {Rich}, {Sanders}, {Schinnerer},
  {Schminovich}, {Shopbell}, {Taniguchi}, \& {Tyson}}]{Scoville2007a}
{Scoville}, N., {Aussel}, H., {Brusa}, M., {et~al.} 2007, \apjs, 172, 1

\bibitem[{{Shapley}(2011)}]{Shapley2011a}
{Shapley}, A.~E. 2011, \araa, 49, 525

\bibitem[{{Sirianni} {et~al.}(2005){Sirianni}, {Jee}, {Ben{\'{\i}}tez},
  {Blakeslee}, {Martel}, {Meurer}, {Clampin}, {De Marchi}, {Ford}, {Gilliland},
  {Hartig}, {Illingworth}, {Mack}, \& {McCann}}]{Sirianni2005a}
{Sirianni}, M., {Jee}, M.~J., {Ben{\'{\i}}tez}, N., {et~al.} 2005, \pasp, 117,
  1049

\bibitem[{{Stanek} {et~al.}(2006){Stanek}, {Gnedin}, {Beacom}, {Gould},
  {Johnson}, {Kollmeier}, {Modjaz}, {Pinsonneault}, {Pogge}, \&
  {Weinberg}}]{Stanek2006a}
{Stanek}, K.~Z., {Gnedin}, O.~Y., {Beacom}, J.~F., {et~al.} 2006, \actaa, 56,
  333

\bibitem[{{Steidel} {et~al.}(1996){Steidel}, {Giavalisco}, {Pettini},
  {Dickinson}, \& {Adelberger}}]{Steidel1996a}
{Steidel}, C.~C., {Giavalisco}, M., {Pettini}, M., {Dickinson}, M., \&
  {Adelberger}, K.~L. 1996, \apj, 462, L17

\bibitem[{{Tanvir} {et~al.}(2004){Tanvir}, {Barnard}, {Blain}, {Fruchter},
  {Kouveliotou}, {Natarajan}, {Ramirez-Ruiz}, {Rol}, {Smith}, {Tilanus}, \&
  {Wijers}}]{Tanvir2004a}
{Tanvir}, N.~R., {Barnard}, V.~E., {Blain}, A.~W., {et~al.} 2004, \mnras, 352,
  1073

\bibitem[{{Tanvir} {et~al.}(2009){Tanvir}, {Fox}, {Levan}, {Berger},
  {Wiersema}, {Fynbo}, {Cucchiara}, {Kr{\"u}hler}, {Gehrels}, {Bloom},
  {Greiner}, {Evans}, {Rol}, {Olivares}, {Hjorth}, {Jakobsson}, {Farihi},
  {Willingale}, {Starling}, {Cenko}, {Perley}, {Maund}, {Duke}, {Wijers},
  {Adamson}, {Allan}, {Bremer}, {Burrows}, {Castro-Tirado}, {Cavanagh}, {de
  Ugarte Postigo}, {Dopita}, {Fatkhullin}, {Fruchter}, {Foley}, {Gorosabel},
  {Kennea}, {Kerr}, {Klose}, {Krimm}, {Komarova}, {Kulkarni}, {Moskvitin},
  {Mundell}, {Naylor}, {Page}, {Penprase}, {Perri}, {Podsiadlowski}, {Roth},
  {Rutledge}, {Sakamoto}, {Schady}, {Schmidt}, {Soderberg}, {Sollerman},
  {Stephens}, {Stratta}, {Ukwatta}, {Watson}, {Westra}, {Wold}, \&
  {Wolf}}]{Tanvir2009a}
{Tanvir}, N.~R., {Fox}, D.~B., {Levan}, A.~J., {et~al.} 2009, \nat, 461, 1254

\bibitem[{{Tanvir} {et~al.}(2012){Tanvir}, {Levan}, {Fruchter}, {Fynbo},
  {Hjorth}, {Wiersema}, {Bremer}, {Rhoads}, {Jakobsson}, {O'Brien}, {Stanway},
  {Bersier}, {Natarajan}, {Greiner}, {Watson}, {Castro-Tirado}, {Wijers},
  {Starling}, {Misra}, {Graham}, \& {Kouveliotou}}]{Tanvir2012a}
{Tanvir}, N.~R., {Levan}, A.~J., {Fruchter}, A.~S., {et~al.} 2012, \apj, 754,
  46

\bibitem[{{Tody}(1986)}]{Tody1986a}
{Tody}, D. 1986, in Society of Photo-Optical Instrumentation Engineers (SPIE)
  Conference Series, Vol. 627, Instrumentation in astronomy VI, ed. D.~L.
  {Crawford}, 733

\bibitem[{{Tremonti} {et~al.}(2004){Tremonti}, {Heckman}, {Kauffmann},
  {Brinchmann}, {Charlot}, {White}, {Seibert}, {Peng}, {Schlegel}, {Uomoto},
  {Fukugita}, \& {Brinkmann}}]{Tremonti2004a}
{Tremonti}, C.~A., {Heckman}, T.~M., {Kauffmann}, G., {et~al.} 2004, \apj, 613,
  898

\bibitem[{{Trenti} {et~al.}(2015){Trenti}, {Perna}, \& {Jimenez}}]{Trenti2014a}
{Trenti}, M., {Perna}, R., \& {Jimenez}, R. 2015, \apj, 802, 103

\bibitem[{{Vergani} {et~al.}(2014){Vergani}, {Salvaterra}, {Japelj}, {Le
  Floc'h}, {D'Avanzo}, {Fernandez-Soto}, {Kr{\"u}hler}, {Melandri}, {Boissier},
  {Covino}, {Puech}, {Greiner}, {Hunt}, {Perley}, {Petitjean}, {Hammer},
  {Levan}, {Mannucci}, {Campana}, {Flores}, {Gomboc}, \&
  {Tagliaferri}}]{Vergani2014a}
{Vergani}, S.~D., {Salvaterra}, R., {Japelj}, J., {et~al.} 2014, \aap,
  submitted, arXiv:1409.7064

\bibitem[{{Wolfe} {et~al.}(2005){Wolfe}, {Gawiser}, \&
  {Prochaska}}]{Wolfe2005a}
{Wolfe}, A.~M., {Gawiser}, E., \& {Prochaska}, J.~X. 2005, \araa, 43, 861

\bibitem[{{Woosley}(2011)}]{Woosley2011a}
{Woosley}, S.~E. 2011, arXiv:1105.4193

\bibitem[{{Yolda{\c s}} {et~al.}(2008){Yolda{\c s}}, {Kr{\"u}hler}, {Greiner},
  {Yolda{\c s}}, {Clemens}, {Szokoly}, {Primak}, \& {Klose}}]{Yoldas2008a}
{Yolda{\c s}}, A.~K., {Kr{\"u}hler}, T., {Greiner}, J., {et~al.} 2008, in
  American Institute of Physics Conference Series, Vol. 1000, American
  Institute of Physics Conference Series, ed. M.~{Galassi}, D.~{Palmer}, \&
  E.~{Fenimore}, 227

\end{thebibliography}

\end{document}